\documentclass[11pt]{article}

\usepackage{graphics}
\usepackage{amsmath, amssymb}
\usepackage{enumerate}
\usepackage{booktabs}
\usepackage{setspace}
\usepackage{appendix}
\usepackage{tikz-cd}
\usepackage[vcentermath]{youngtab}

\newcommand{\tabincell}[2]{\begin{tabular}{@{}#1@{}}#2\end{tabular}}

\numberwithin{equation}{section}

\allowdisplaybreaks

\usepackage{graphicx}
\newcommand{\p}{\partial}
\newcommand{\e}{\epsilon}

\renewcommand{\P}{\mathbb{P}^1}

\newcommand{\QQ}{\mathbb{Q}}

\newcommand{\CC}{\mathbb{C}}
\newcommand{\ZZ}{\mathbb{Z}}

\newcommand{\F}{\mathcal{F}}
\newcommand{\U}{\mathcal{U}}

\newcommand{\D}{\mathcal{D}}
\newcommand{\R}{\mathcal{R}}

\newcommand{\zcp}{Z_{\mathbb P^1}}

\newcommand{\li}{{\rm Li}}

\newcommand{\fcp}{\mathcal{F}^{\mathbb P^1}}

\newcommand{\beq}{\begin{equation}}
\newcommand{\eeq}{\end{equation}}

\newcommand{\nn}{\nonumber}

\newcommand{\llangle}{\langle\!\langle}
\newcommand{\rrangle}{\rangle\!\rangle}

\newtheorem{dfn}{Definition}[section]
\newtheorem{lem}[dfn]{Lemma}
\newtheorem{prp}[dfn]{Proposition}

\newtheorem{rmk}[dfn]{Remark}

\newtheorem{cnj}[dfn]{Conjecture}

\newenvironment{prf}{\noindent {\it Proof} \ }{\hfill $\Box$}

\begin{document}

\title{On Equivariant Gromov--Witten Invariants of Resolved Conifold with Diagonal and Anti-Diagonal Actions} 
\author{Si-Qi Liu, \quad Di Yang, \quad Youjin Zhang, \quad Chunhui Zhou}
\date{}
\maketitle

\begin{abstract}
We propose two conjectural relationships between the
 equivariant Gromov-Witten invariants of the resolved conifold under diagonal and anti-diagonal actions  
 and the Gromov-Witten invariants of $\P$, and verify their validity in genus zero approximation. We also provide evidences to support the validity of these relationships in genus one and genus two. 
\end{abstract}

\tableofcontents

\section{Introduction}
Let $g,d,n$ be non-negative integers, and
\[
\langle\tau_{k_1}(\phi_{\alpha_1})\cdots \tau_{k_n}(\phi_{\alpha_n})\rangle_{g,d}^{\P}
\]
be the Gromov--Witten (GW) invariants of~$\P$ of genus $g$ and degree~$d$, 
where $\alpha_1,\dots,\alpha_n\in\{1,2\}$ and the integers $k_1,\dots,k_n\geq0$.
The generating function 
\beq
\mathcal{F}^{\mathbb{P}^1}({\bf s};q;\e) := \sum_{g\geq0} \e^{2g-2} \sum_{n,d\geq0}
\frac{s^{\alpha_1,k_1}\cdots s^{\alpha_n,k_n}}{n!} q^d  
\langle\tau_{k_1}(\phi_{\alpha_1})\cdots \tau_{k_n}(\phi_{\alpha_n})\rangle_{g,d}^{\P}
\eeq
of these numbers is called the {\it free energy} of the GW invariants of $\P$.
Here ${\bf s}:=(s^{\alpha,k})_{\alpha=1,2;k\geq0}$ is an infinite vector of indeterminates, $\e$ is an indeterminate called 
the {\it string coupling constant}, and
summation over repeated upper and lower Greek indices is assumed.
For the definition of GW invariants see~\cite{BF, KM, LT, RT1, W}.
The free energy has the following form of genus expansion:
\beq
 \mathcal{F}^{\mathbb{P}^1}({\bf s};q;\e) =: \sum_{g\ge 0} \e^{2g-2} \fcp_g({\bf s};q),
\eeq
where $\fcp_g({\bf s};q)$ is called the {\it genus~$g$ free energy}.
The exponential of the free energy 
\beq\label{defzcp}
Z_{\P}({\bf s};q;\e):=e^{\mathcal{F}^{\mathbb{P}^1}({\bf s};q;\e)}
\eeq
is called the {\it partition function} of the GW invariants of~$\P$.

The restriction of the genus zero free energy to the small phase space 
yields the potential 
\beq\label{fp1nov}
F^{\mathbb{P}^1}=\frac12 (v^1)^2 v^2 + q e^{v^2}
\eeq 
of a two-dimensional Frobenius manifold~\cite{Du1}, which is called the $\P$-Frobenius manifold denoted by~$M_{\P}$.
We know from \cite{Du1} that the genus zero free energy $\mathcal{F}_0^{\mathbb{P}^1}({\bf s};q)$ 
can be reconstructed in terms of the so-called topological solution to the Principal Hierarchy of the Frobenius manifold~$M_{\P}$
(see Section~\ref{section2} for the definition of the Principal Hierarchy of a Frobenius manifold).
Moreover, the higher genus   
free energies $\fcp_g({\bf s};q)$, $g\ge1$, can be obtained by solving 
the {\it loop equation} of the Frobenius manifold \cite{DZ-norm}, and the partition function $Z_{\P}({\bf s};q;\e)$ is a particular 
tau-function of the extended Toda hierarchy \cite{DZ1} 
(cf.~also \cite{CDZ, Getz, OP, OP1, Z}).

Now let us consider the equivariant GW invariants of the target
\[
X= \left(\mathcal O_{\mathbb P^1}(-1)\oplus \mathcal O_{\mathbb P^1}(-1)\right)^{\circlearrowleft \mathcal T},
\]
where $\mathcal T\simeq \CC^*$ is the torus action with two characters $(\kappa_1,\kappa_2)$ on the fibers.
Denote by
\[\langle\tau_{k_1}(\phi_{\alpha_1})\cdots \tau_{k_n}(\phi_{\alpha_n})\rangle_{g,d}^{X,\rm di/ad}\]
 the genus $g$ and degree $d$ equivariant GW invariants of~$X$ under diagonal or anti-diagonal action, 
 which corresponds respectively to the case with $\kappa_1=\kappa_2=1$ or to the case $\kappa_1=-\kappa_2=1$
  (cf.~\cite{BP}).
Define the {\it partition function} of the equivariant GW invariants of~$X$ with diagonal or anti-diagonal action by  
\beq\label{defzloc}
Z_{X,\rm di/ad}({\bf t};q;\e)=e^{\F^{X,\rm di/ad}({\bf t};q;\e)},
\eeq
where ${\bf t}=(t^{\alpha,k})_{\alpha=1,2;k\geq0}$, and
\beq\label{defz}
\F^{X,\rm di/ad}({\bf t};q;\e):=\sum_{g\geq0}\e^{2g-2}\sum_{d\geq0}q^d 
\sum_{n\geq0}\frac{t^{\alpha_1,k_1}\cdots t^{\alpha_n,k_n}}{n!}
\langle\tau_{k_1}(\phi_{\alpha_1})\cdots \tau_{k_n}(\phi_{\alpha_n})\rangle_{g,d}^{X,\rm di/ad}
\eeq
is the {\it free energy} of the equivariant GW invariants of~$X$ with diagonal or anti-diagonal action. 
The restrictions of the genus zero free energies to the small phase space yield the primary genus zero free energies~\cite{Bri, BCR} 
\begin{align}
& F^{X, \rm di} = \frac12 (u^1)^2 u^2+\frac13 (u^1)^3+{\rm Li}_3\Bigl(q e^{u^2}\Bigr), \label{FhatNov}\\
& F^{X, \rm ad} = \frac12 (u^1)^2 u^2 - {\rm Li}_3\Bigl(qe^{u^2}\Bigr),\label{FalhatNov}
\end{align}
which serve as potentials of two Frobenius manifolds.
Here ${\rm Li}_3$ denotes a special polylogarithmic function, i.e., 
\beq
{\rm Li}_k(z):=\sum_{n\geq1} \frac{z^n}{n^k},\quad k\in\ZZ.
\eeq
We denote these two Frobenius manifolds by $M_{X, \rm di}$ and  $M_{X, \rm ad}$ respectively, and  
we note that these two Frobenius manifolds do not possess Euler vector fields.
Before proceeding,
we introduce the following notations that will be used later:
\begin{align}
&b_{1,n}^{\alpha,m,\rm di}=\frac{n^m}{n!}\left(\delta^{\alpha,1}
+2m\delta^{\alpha,2}\right),\quad
b_{2,n}^{\alpha,m,\rm di}=\frac{(n+1)^m}{n!}\delta^{\alpha,2},
\label{constb}
\\
&b_{1,n}^{\alpha,m,\rm ad}=\frac{(\sqrt{-1})^{n-1} n^m}{n!}\left(\delta^{\alpha,1}
+(2m+1)\delta^{\alpha,2}\right),\quad
b_{2,n}^{\alpha,m,\rm ad}=\frac{(\sqrt{-1})^n (n+1)^m}{n!}\delta^{\alpha,2},
\label{constb1}
\end{align}
where $\alpha=1,2$, and $m,n$ are non-negative integers.

Let us proceed and propose two conjectural relationships 
between the equivariant GW invariants of the resolved conifold under diagonal and anti-diagonal actions  
 and the GW invariants of~$\P$ via the theory of Frobenius manifold.

For the diagonal case, the conjectural relationship can be interpreted by observing the {\it almost duality} 
between the Frobenius manifolds $M_{\P}$ and $M_{X, \rm di}$.
From the definition of almost duality introduced by Dubrovin in~\cite{Du2} it follows that  
the potential of the almost dual $\widehat M_{\P}$ of~$M_{\P}$ coincides with \eqref{FhatNov} if we choose the 
flat coordinates of~$\widehat M_{\P}$ as follows:
\begin{align}
&u^1=\log\left(v^1+\sqrt{(v^1)^2-4qe^{v^2}}\right)-\log{2},\label{zh-3}\\
&u^2=v^2-2\log\left(v^1+\sqrt{(v^1)^2-4qe^{v^2}}\right)+2\log 2.\label{zh-4}
\end{align}
For the definition of~$\widehat M_{\P}$ see Section~\ref{section2}. 
It then follows from a general principle (cf.~Lemma~\eqref{almostdualityflows} of Section~\ref{section2}) that tau-symmetric hamiltonian densities 
of the Principal Hierarchy of~$\widehat M_{\P}$ are linear combinations  
of those of the Principal Hierarchy of $\widehat M_{\P}$.
Motivated by the 
Hodge-GUE correspondence~\cite{DLYZ, DY2} (cf.~also~\cite{DLYZ0, LYZZ}), 
we then propose an explicit conjectural 
relationship
between $Z_{X,\rm di}({\bf t};q;\e)$ and $Z_{\P}({\bf s};q;\e)$.
\begin{cnj}\label{cnjmain}
The identity 
\beq\label{zcpzloc}
\zcp({\bf s}; q;\epsilon)=\exp\left(\frac{A_{\rm di}({\bf s})}{\epsilon^2}\right) 
Z_{X,{\rm di}}({\bf t}_{\rm di}({\bf s});q; \epsilon)
\eeq
holds true in 
\beq\label{defv}
V_{\rm di;\e}:=\CC((\epsilon)) \otimes \CC[[s^{1,0}-1,s^{2,0},s^{1,1},s^{2,1},\dots;q]].
\eeq
Here $A_{\rm di}({\bf s})$ and ${\bf t}_{\rm di}({\bf s})$ are defined by 
\begin{align}
&A_{\rm di}({\bf s}):=
\sum_{k,\ell\geq0}\frac{s^{1,k}  s^{2,\ell}}{(k+\ell+1)k!\ell!} 
-\sum_{k\geq 0}\frac{s^{2,k}}{(k+2)k!},\\
&t_{\rm di}^{\alpha,k}({\bf s}):=\sum_{\ell\geq0} b^{\alpha,k,\rm di}_{\beta,\ell} 
s^{\beta,\ell}-b_{1,1}^{\alpha,k,\rm di}+\delta^{\alpha,1}\delta^{k,1},
\quad  \alpha=1,2, \, k\geq0.  \label{ts}
\end{align}
\end{cnj}

Let us now consider the anti-diagonal case. It was found in~\cite{Bri, BCR} (see also~\cite{Du3, Du4}) that the function $F^{X, \rm ad}$ is the potential of the almost dual $\widehat M_{\rm AL}$ of the Frobenius manifold $M_{\rm AL}$ associated with the Ablowitz--Ladik (AL) hierarchy. The potential of $M_{\rm AL}$ is given by
\beq
F_{\rm AL} = \frac12 (t^1)^2 t^2 + q \, t^1 e^{t^2} + \frac12 (t^1)^2 \log t^1,
\eeq
where $t^1,t^2$ are related to $u^1,u^2$  by 
\beq\label{zh-5}
t^1 = \sqrt{-1} e^{u^1} (1- q e^{u^2})  , \quad t^2 = u^1+u^2.
\eeq
In~\cite{Bri} Brini conjectured that the generating function of the equivariant GW invariants of $X$
with anti-diagonal action is the logarithm of a tau-function of the AL hierarchy.
The main motivation that leads to his conjecture is the fact
that the quasi-trivial transformation of the AL equation 
gives the genus expansion of the free energy of $X$ with anti-diagonal action up to genus one,
as well as some evidences of validity of his conjecture at genus two.
According to the general approach of Dubrovin--Zhang (DZ) relating GW invariants with integrable hierarchies,
the anti-diagonal equivariant GW invariants of $X$ 
should be related with 
the topological deformation of the Principal Hierarchy of~$\widehat M_{\rm AL}$. 
More precisely, the partition function of the anti-diagonal equivariant GW invariants of $X$
should be a tau-function of the DZ hierarchy of~$\widehat M_{\rm AL}$,
and the genus expansion of the free energy of $X$ with anti-diagonal action should be given by the quasi-trivial transformation relating the Principal Hierarchy of $\widehat M_{\rm AL}$ with this DZ hierarchy.
In Brini's approach~\cite{Bri},
he used the quasi-trivial transformation of the AL equation to arrive at the conjectural genus expansion of the free energy of $X$ with anti-diagonal action.
The reason, to our understanding, is
that the AL equation should be obtained by the Miura transformation of a certain non-trivial infinite
linear combination of the flows of the DZ hierarchy of $\widehat M_{\rm AL}$ (cf.~Lemma~\ref{almostdualityflows}),
so the flow given by the AL equation commutes with the flows of the DZ hierarchy of $\widehat M_{\rm AL}$,
thus the quasi-trivial transformation of the AL hierarchy coincides with 
that of the DZ hierarchy of $\widehat M_{\rm AL}$.  
On the other hand, the Frobenius manifold 
$M_{\rm AL}$ is related to the Frobenius manifold $M_{\P}$ by a Legendre-type transformation (see Appendix B of \cite{Du1}):
\[v^1=\frac{\p^2 F_{\rm AL}}{\p t^1\p t^2},\quad
v^2=\frac{\p^2 F_{\rm AL}}{\p t^1\p t^1},\]
and 
\[\frac{\p^2 F^{\mathbb{P}^1}}{\p v^\alpha\p v^\beta}=\frac{\p^2 F_{\rm AL}}{\p t^\alpha\p t^\beta},\quad \alpha, \beta=1,2.\]
Since the tau-functions of the DZ hierarchies of $M_{\rm AL}$ and $M_{\P}$ are equivalent in a certain sense,
(cf.~\cite{CDZ}; see also~\cite{V2}),
it is then more convenient for us to use the partition function of $M_{\P}$
to give an explicit relationship between the equivariant GW invariants of $X$ with anti-diagonal action and the extended Toda hierarchy
which governs the GW invariants of $\P$~\cite{DZ1, Getz, Z}.
Motivated by Brini's conjecture, by the Hodge-GUE correspondence~\cite{DLYZ}, and by the above-mentioned Legendre-type transformation between $M_{\rm AL}$ and $M_{\P}$,
we are to propose an explicit conjectural 
relationship between $Z_{X,\rm ad}({\bf t};q;\e)$ and $Z_{\P}({\bf s};q;\e)$ in Conjecture~\ref{cnjmain1}.

\begin{cnj}\label{cnjmain1}
The identity 
\beq\label{zcpzloc1}
\zcp({\bf s};q;\epsilon)
=\exp\left(\frac{A_{\rm ad}({\bf s})}{\epsilon^2}\right) 
Z_{X,\rm ad}({\bf t}_{\rm ad}({\bf s});q; \sqrt{-1}\epsilon)
\eeq
holds true in 
\beq\label{defv1}
V_{\rm ad;\e}:=\CC((\epsilon)) \otimes \CC[[s^{1,0}-\sqrt{-1},s^{2,0},s^{1,1},s^{2,1},\dots;q]].
\eeq
Here $A_{\rm ad}({\bf s})$ and $t_{\rm ad}({\bf s})$ are defined by 
\begin{align}
&A_{\rm ad}({\bf s}):=
\sum_{k,\ell\geq0}\frac{(\sqrt{-1})^{k+\ell+1}}{(k+\ell+1)k!l!} s^{1,k}  s^{2,\ell}
+\sum_{k\geq 0}\frac{(\sqrt{-1})^k}{(k+2)k!}s^{2,k},\\
&t_{\rm ad}^{\alpha,k}({\bf s})
:=\sum_{\ell\geq0} b^{\alpha,k,\rm ad}_{\beta,\ell}s^{\beta,\ell}
-b_{1,1}^{\alpha,k,\rm ad}+\delta^{\alpha,1}\delta^{k,1},
\quad \alpha=1,2, \, k\geq0.\label{ts1}
\end{align}
\end{cnj}

\begin{rmk}
Since we  know that $\F^{\P}=\log Z_{\P}({\bf s};q;\e)$ belongs to the space
\[\e^{-2}\CC[[\e^2]]\otimes\CC[s^{1,0}][[s^{2,0},s^{1,1},s^{2,1},\dots;q]],\]
the conjectural identities~\eqref{zcpzloc} and~\eqref{zcpzloc1} imply that the logarithms of the right-hand sides belong to this space.
\end{rmk}

From the definitions~\eqref{defzcp} and~\eqref{defzloc} we see that 
the conjectural identity~\eqref{zcpzloc} is equivalent to the following identities:
\begin{align}
\fcp_g({\bf s};q)
=\F^{X,{\rm di}}_g({\bf t}_{\rm di}({\bf s});q)
+A_{\rm di}({\bf s}) \delta_{g,0},
\quad g\geq0,\label{fgfg} 
\end{align}
where $\F^{X,\rm di}_g({\bf t};q)$ denotes the genus~$g$ free energy 
of the equivariant GW invariants of~$X$ with diagonal action. 
Similarly, the conjectural identity~\eqref{zcpzloc1} is equivalent to
\beq \label{fgfg1}
\fcp_g({\bf s};q)
=(-1)^{g-1} \F^{X,{\rm ad}}_g({\bf t}_{\rm ad}({\bf s});q)\
+A_{\rm ad}({\bf s}) \delta_{g,0},
\quad g\geq0,
\eeq
where $\F^{X,\rm ad}_g({\bf t};q)$ denotes the genus~$g$ free energy 
of the equivariant GW invariants of~$X$ with anti-diagonal action.

For $g=0$, there are explicit expressions for both sides of the conjectural identities~\eqref{fgfg} and~\eqref{fgfg1}. 
We will verify the validity of these identities in Section~\ref{g0verification}.

For $g\geq1$, the verification is more involved because, as far as we know,
there are no efficient algorithms to compute the right-hand sides of \eqref{fgfg} and~\eqref{fgfg1}.
However, since the free energies  $\F^{X,\rm di}_g({\bf t};q)$,  $\F^{X,\rm ad}_g({\bf t};q)$ and $\fcp_g({\bf s};q)$
can be represented in terms of the so-called jet variables~\cite{DW,DZ-norm,DZ1,EYY},
from the conjectural 
identities \eqref{fgfg} and \eqref{fgfg1} we can determine the functions $\F^{X,\rm di}_g({\bf t};q)$,  $\F^{X,\rm ad}_g({\bf t};q)$ in terms of the function $\fcp_g({\bf s};q)$. 
We are then to give supports of the validity of the identities \eqref{fgfg} and~\eqref{fgfg1} for $g=1, 2$ by
proving that the functions $\F^{X,\rm ad}_g({\bf t};q)$ and $\F^{X,\rm di}_g({\bf t};q)$ determined by these identities satisfy the genus one and genus two topological recursion relations given in~\cite{BP, G, G2}.

The paper is organized as follows.
In Section~\ref{section2} we  recall the definition of the Principal Hierarchy of a Frobenius manifold. 
In Section~\ref{g0verification}, 
we give a proof of the identities~\eqref{fgfg} and~\eqref{fgfg1} for $g=0$.
In Section~\ref{g12},
we derive the expressions of the free energies 
$\F^{X,\rm di/ad}_g({\bf t};q)$ in terms of $\fcp_g({\bf s};q)$ for $g=1, 2$, and show that they satisfy the genus one and genus two topological recursion relations.
In Section~\ref{section5},
we give some further remarks.

\paragraph{Acknowledgements} 
We would like to thank Paolo Lorenzoni for helpful discussions on Lemma~\ref{almostdualityflows}.
This work is partially supported by NSFC No.\,12171268, No.\,11725104 and No.\,12061131014.

\section{The Principal Hierarchy of a Frobenius manifold}\label{section2}

In this section, we recall the definitions of the Principal Hierarchy and almost duality 
of a Frobenius manifold (for details see~\cite{Du1, Du2, DZ-norm}).

Let $(M, \cdot, \eta, e, E)$ be an $n$-dimensional Frobenius manifold of charge~$d$, where $\cdot$ denotes the operation of multiplication on the tangent spaces of $M$, $\eta$ is the invariant flat metric, $e$ is the unit vector field and $E$ is the Euler vector field.
Fix a system of flat coordinates $v^1,\dots, v^n$ of the metric $\eta$ such that $e=\frac{\p}{\p v^1}$. 
In these coordinates we have 
\[
\frac{\p}{\p v^\alpha}\cdot \frac{\p}{\p v^\beta}
=c_{\alpha\beta}^\gamma(v) \frac{\p}{\p v^\gamma}.
\]
Here and below, free Greek indices always take $1,\dots,n$. 
Denote 
\[
\eta_{\alpha\beta}=\eta\left(\frac{\p}{\p v^\alpha},\frac{\p}{\p v^\beta}\right),\quad
c_{\alpha\beta\gamma}(v):=\eta_{\alpha\sigma}c^\sigma_{\beta\gamma},\]
then the potential $F(v)=F(v^1,\dots, v^n)$ of the Frobenius manifold $M$ is defined by 
\[
c_{\alpha\beta\gamma}=\frac{\p^3 F(v)}{\p v^\alpha \p v^\beta \p v^\gamma}.
\]

An important geometric object of a Frobenius manifold is the deformed flat connection 
$\widetilde \nabla$ defined by
\beq
\widetilde \nabla_a b=\nabla_a b+z a\cdot b,\quad a,b\in TM.
\eeq
One can find a system of deformed flat coordinates of the form
\[(\tilde{v}_1(v; z),\dots,\tilde{v}_n(v;z))=(\theta_1(v;z),\dots,\theta_n(v;z)) z^\mu z^R\]
which satisfy the equations
\beq\label{zh-1}
\widetilde \nabla d\tilde{v}_\alpha(v;z)=0.
\eeq
Here the constant matrices $\mu=\rm{diag}(\mu_1,\dots,\mu_n)$, $R=R_1+\dots+R_m$ are the monodromy data of the Frobenius manifold
at $z=0$, and $\theta_\alpha(v;z)$ have the expressions
\beq\label{thetaz}
\theta_\alpha(v;z)=\sum_{k\geq0}\theta_{\alpha,k}(v)z^k.
\eeq
In terms of the functions $\theta_{\alpha,p}$, the equations \eqref{zh-1} can be rewritten as
\beq\label{thetarecursion}
\frac{\p^2 \theta_{\gamma,p+1}}{\p v^\alpha \p v^\beta}
=c_{\alpha\beta}^\xi \frac{\p \theta_{\gamma,p}}{\p v^\xi},\quad  p\ge 0.
\eeq
We can also require that these deformed flat coordinates satisfy the following normalization conditions:
\begin{align}
&\theta_{\alpha}(v;0)=\eta_{\alpha\beta}v^\beta,\quad
\frac{\p\theta_\alpha(v;z)}{\p v^1}=z\theta_\alpha(v;z)+\eta_{1\alpha},
\label{falpha1}\\
&\frac{\p \theta_\alpha(v;z)}{\p v^\gamma}\eta^{\gamma\sigma} \frac{\p\theta_\beta(v;-z)}{\p v^\sigma}
=\eta_{\alpha\beta},\quad \rm{with}\ (\eta^{\alpha\beta}):=(\eta_{\alpha\beta})^{-1},
\label{thetacond}
\end{align}
and the quasi-homogeneity condition
\beq\label{zh-3b}
\p_E\theta_{\alpha,p}(v)=\left(p+\frac{2-d}2+\mu_\alpha\right)\theta_{\alpha,p}(v)+\sum_{k=1}^p\theta_{\xi,p-k}(v)\left(R_k\right)^\xi_\alpha+\rm{constant}.
\eeq

For the Frobenius manifold $M_{\P}$, we have
\[\mu_1=-\mu_2=-\frac12,\quad R=R_1=\begin{pmatrix} 0 &0\\ 2 &0\end{pmatrix}.\]
From the conditions \eqref{falpha1}--\eqref{zh-3b} it follows that the functions $\theta_{\alpha}(v;z)$ can be chosen as \cite{DZ-norm, DZ1} 
\begin{align}
\theta^{\P}_1(v;q;z)&=-2 e^{z v^1}\left(K_0\!\left(2z\sqrt{q e^{v^2}}\right)+(\log z+\gamma+\frac12\log q)I_0\!\left(2z\sqrt{q e^{v^2}}\right)\right) \nn \\
&=-2e^{zv^1}\sum_{k\geq0}\left(\gamma-\frac12 v^2+\psi(k+1)\right)q^k e^{k v^2} \frac{z^{2k}}{(k!)^2},
\label{p1theta1}\\
\theta^{\P}_2(v;q;z)&=z^{-1}e^{z v^1}I_0\!\left(2z\sqrt{q e^{v^2}}\right)-z^{-1} \nn \\
&=z^{-1}\left(\sum_{k\geq0}q^ke^{kv^2+zv^1}\frac{z^{2k}}{(k!)^2}-1\right),\label{p1theta2}
\end{align}
where $I_0$ and $K_0$ are the modified Bessel functions of the first and second kinds, 
$\gamma$ is the Euler's constant, 
$\psi(z)$ is the digamma function, 
and we also indicate the dependence of the functions $\theta_{\alpha, p}$ on the parameter $q$.

The \emph{Principal Hierarchy} of $M$ is defined as the following system of evolutionary Hamiltonian PDEs of hydrodynamic type: 
\beq\label{hierarchy}
\frac{\p v^\alpha}{\p s^{\beta,p}}=\eta^{\alpha \gamma}\p_x\frac{\p \theta_{\beta,p+1}}{\p v^\gamma},
\quad p\geq0.
\eeq
Since the flow $\frac{\p}{\p s^{1,0}}$ reads
\[
\frac{\p v^{\alpha}}{\p s^{1,0}}=\frac{\p v^\alpha}{\p x},
\]
we identify the variable $s^{1,0}$ with the spatial variable~$x$. 
It is shown in~\cite{Du1} that the Principal Hierarchy is integrable. 
In particular, the 
Hamiltonian densities $\theta_{\alpha,p}$, $p\geq0$, are conserved densities 
for each flow of the Principal Hierarchy, and therefore they satisfy the following second order linear PDEs~\cite{DZ-norm}:
\beq\label{pdeham}
c^{\beta}_{\rho\alpha} \frac{\p^2 h}{\p v^{\beta} \p v^\sigma} = c^{\beta}_{\rho\sigma} \frac{\p^2 h}{\p v^{\beta} \p v^\alpha}.
\eeq

The \emph{topological solution} $v_{\rm top}({\bf s})$ to the Principal Hierarchy
is specified by the initial condition
\beq
v_{\rm top}^\alpha\mid_{s^{\beta,k}=\eta^{\beta1}\delta^{k,0}x}=\delta^{\alpha,1} x.
\eeq
It can be solved uniquely from the following genus zero Euler--Lagrange equation:
\beq\label{eleq}
\sum_{p\geq0}\tilde s^{\alpha,p}\frac{\p \theta_{\alpha,p}(v)}{\p v^\beta}=0,
\eeq
where $\tilde s^{\alpha,p}=s^{\alpha,p}-\delta^{\alpha,1}\delta^{p,1}$.

We call the following power series 
\beq\label{f0omega}
\F_0=\frac12\sum_{k,\ell\geq0}\tilde s^{\alpha,k}\tilde s^{\beta,\ell} 
\Omega_{\alpha,k;\beta,\ell}\left(v_{\rm top}({\bf s})\right)\in \CC[[{\bf s}]]
\eeq
 the genus zero free energy of the Frobenius manifold, where the functions $\Omega_{\alpha,k;\beta,l}(v)$ are defined by 
\beq\label{defomega}
\sum_{k,\ell\geq0}\Omega_{\alpha,k;\beta,\ell}(v)z^k w^\ell
:=\frac{\eta(\nabla\theta_\alpha(v;z),\nabla\theta_\beta(v;w))
-\eta_{\alpha\beta}}{z+w}.
\eeq

Now let us recall the definition of the {\it almost dual} $\widehat{M}$ of a Frobenius manifold $M$ introduced in \cite{Du2}. Let $g$ be the intersection form of $M$ defined by 
\beq\label{metric2}
g(v)(a,b):=i_E (a\cdot b) ,\quad a,b\in T_v^*M,
\eeq
where the multiplication operation on the cotangent space is induced from the one defined on tangent space by the metric $\eta$. Then we can introduce a Frobenius manifold structure on 
\[\widehat M=M\setminus \{v \in M\mid g(v)\,{\rm is\ degenerate\ on}\, T^*_vM\}\]
by defining a new product
\beq
a*b=E^{-1}\cdot a\cdot b,\quad a,b\in T_vM.
\eeq
The flat metric and the unit vector field of $\widehat M$ are given by $\hat\eta=g^{-1}$ and $E$
respectively. We note that on the almost dual $\widehat M$ there is no Euler vector field in general.
Choose a system of local flat coordinates $u^1,\dots,u^n$ for the metric~$\hat\eta$, 
then there exists a function $\widehat F$ satisfying the equations
\beq\label{defstrucalmostdual}
\frac{\p^3 \widehat F}{\p u^\alpha \p u^\beta \p u^\gamma}=\hat \eta_{\alpha\xi}
\hat \eta_{\beta\zeta}
\frac{\p u^{\xi}}{\p v^\sigma}\frac{\p u^{\zeta}}{\p v^\rho}
\frac{\p v^\delta}{\p u^\gamma}
c^{\sigma\rho}_\delta.
\eeq
We call $\widehat F$ the {\it almost dual potential}. The structure constants for~$\widehat M$ will be denoted by 
$\widehat c^{\alpha}_{\beta\gamma}$.

The following lemma which connects the conserved densities of the Principal Hierarchies of $M$ and $\widehat M$
will be used later.
\begin{lem}\label{almostdualityflows}
If $h(v)$ is a conserved density for the Principle Hierarchy of~$M$, then $h(v(u))$ is a conserved density of the Principle Hierarchy of~$\widehat M$. 
\end{lem}
\begin{prf}
By using~\eqref{pdeham}, \eqref{defstrucalmostdual} and the fact that $u^1,\dots,u^n$ are the flat coordinates of~$\hat\eta$,
 it can be verified that~$h$ satisfies the following second order PDEs:
\beq\label{feb1}
\widehat c^{\beta}_{\rho\alpha} \frac{\p^2 h(v(u))}{\p u^{\beta} \p u^\sigma} = \widehat c^{\beta}_{\rho\sigma} \frac{\p^2 h(v(u))}{\p u^{\beta} \p u^\alpha}.
\eeq
The lemma is then proved. 
\end{prf}

As we mentioned in the Introduction, the almost dual $\widehat M_{\P}$ of the Frobenius manifold $M_{\P}$ coincides with the Frobenius manifold associated with the resolved conifold $X$ with diagonal action. 
It has the potential $\widehat F^{\P}=F^{X,\rm di}$ given by \eqref{FhatNov}, and the flat coordinates $u^1, u^2$ are given by~\eqref{zh-3},~\eqref{zh-4}.  

Denote $(a)_k:=a(a+1)\cdots(a+k-1)$. Recall that the deformed flat coordinates for $X$ with diagonal action can be chosen as follows:
\begin{align}
& \theta_1^{X,\rm di}(u;q;z)=
 e^{zu^1}\sum_{d\geq0} q^d e^{d u^2}\frac{(-z)_d(-z)_d}{d!^2}\left( u^2+2z^{-1}-2\sum_{k=1}^{d}\frac{z+1}{k(z-k+1)}\right)
-2z^{-1},
\label{tloc1} \\
& \theta_2^{X,\rm di}(u;q;z)=
 z^{-1}e^{z u^1}
\sum_{d\geq0} q^d e^{d u^2}\frac{(-z)_d (-z)_d}{d!^2}
-z^{-1}.
\label{tloc2}
\end{align}
It can be verified that $\theta^{X,\rm di}_\alpha(u;q;z)$ coincide with the calibrations given by Brini in~\cite{Bri}.
Obviously, $\theta^{X,\rm di}_{\alpha,k}(u;q)$ belong to~$\QQ[u^1,u^2,e^{u^2}][[q]]$.
More precisely, they have the form
\begin{align}
& \theta_{1,k}^{X,\rm di} = \sum_{d\geq0} \sum_{0\leq i \leq [k/2]} c_{i,d,k}^{\rm di} \frac{(u^1)^{k-2i}}{(k-2i)!} (qe^{u^2})^d 
+ u^2 \sum_{d\geq0} \sum_{0\leq i \leq [k/2]} a_{i,d,k}^{\rm di} \frac{(u^1)^{k-2i}}{(k-2i)!} (qe^{u^2})^d, \label{311} \\
& \theta_{2,k}^{X,\rm di} = \sum_{d\geq0} \sum_{0\leq i \leq [(k+1)/2]} a_{i,d,k}^{\rm di} \frac{(u^1)^{k+1-2i}}{(k+1-2i)!} (qe^{u^2})^d,  \label{312}
\end{align}
where $a_{i,d,k}^{\rm di}, c_{i,d,k}^{\rm di}\in \QQ$. 
We note that $a_{i,d,k}^{\rm di}, c_{i,d,k}^{\rm di}$ vanish when $d<i$.

We also mentioned in the Introduction that the almost dual $\widehat M_{\rm AL}$ of the Frobenius manifold $M_{\rm AL}$ associated with the Ablowitz-Ladik hierarchy coincides with the Frobenius manifold associated with the resolved conifold $X$ with anti-diagonal action. The almost dual potential is given by \eqref{FalhatNov}, and the relation of the flat coordinates of the Frobenius manifold $\widehat M_{\rm AL}$ with those of $M_{\rm AL}$ is given by~\eqref{zh-5}. 
The deformed flat coordinates can be chosen as follows:
\begin{align}
& \theta_1^{X,\rm ad}(u;q;z)=
e^{z u^1}\sum_{d\geq0} q^d e^{d u^2}  \frac{(z)_d(-z)_d}{d!^2}\left(u^2+2\sum_{k=0}^{d-1}\frac{k+z^2}{(k+1)(k^2-z^2)}\right), 
\label{adtloc1}\\
& \theta_2^{X,\rm ad}(u;q;z)=
z^{-1}e^{z u^1}
\sum_{d\geq0} q^d e^{d u^2}\frac{(z)_d(-z)_d}{d!^2}
-z^{-1}.
\label{adtloc2}
\end{align}
It can also be verified that $\theta^{X,\rm ad}_\alpha(u;q;z)$ coincide with the calibrations given by Brini in~\cite{Bri}.
Obviously, $\theta^{X,\rm ad}_{\alpha,k}(u;q)$ belong to~$\QQ[u^1,u^2,e^{u^2}][[q]]$, and they have the form
\begin{align}
& \theta_{1,k}^{X,\rm ad} = \sum_{d\geq0} \sum_{0\leq i \leq [k/2]} c_{i,d,k}^{\rm ad} \frac{(u^1)^{k-2i}}{(k-2i)!} (qe^{u^2})^d 
+ u^2 \sum_{d\geq0} \sum_{0\leq i \leq [k/2]} a_{i,d,k}^{\rm ad} \frac{(u^1)^{k-2i}}{(k-2i)!} (qe^{u^2})^d, \label{311ad}\\
& \theta_{2,k}^{X,\rm ad} = \sum_{d\geq0} \sum_{0\leq i \leq [(k+1)/2]} a_{i,d,k}^{\rm ad} \frac{(u^1)^{k+1-2i}}{(k+1-2i)!} (qe^{u^2})^d, \label{312ad}
\end{align}
where $a_{i,d,k}^{\rm ad}, c_{i,d,k}^{\rm ad}\in \QQ$.
We note that $a_{i,d,k}^{\rm ad}, c_{i,d,k}^{\rm ad}$ vanish when $d<i$.

In what follows, 
we will use $\Omega^{\P}_{\alpha,k;\beta,\ell}$, 
$\Omega^{X,\rm di}_{\alpha,k;\beta,\ell}$ 
and $\Omega^{X,\rm ad}_{\alpha,k;\beta,\ell}$
to denote the genus zero two-point correlation functions defined by~\eqref{defomega}
for the Frobenius manifolds $M_{\P}$, $\widehat M_{\P}$ and~$\widehat M_{\rm AL}$.
We also denote by $v_{\rm top}({\bf s};q)$,
$u_{\rm top,di}({\bf t};q)$ 
and $u_{\rm top,ad}({\bf t};q)$
the topological solutions to the Principal Hierarchies of these three Frobenius manifolds.

\section{Verification for the genus zero case}\label{g0verification}

In this section, we will verify the genus zero part of Conjecture~\ref{cnjmain}.

\subsection{The diagonal case}

Let us first prove the following lemma.

\begin{lem}\label{lemvu}
Under the coordinate transformation
\beq\label{vu}
v^1
=e^{u^1}\left(1+q e^{u^2}\right),\quad
v^2
=2 u^1+u^2,
\eeq
the following identities hold true:
\beq\label{thetatheta}
\theta^{\P}_{\alpha,k}(v; q)
= \sum_{\ell\geq0}
b_{\alpha,k}^{\beta,\ell,{\rm di}}
\theta^{X,{\rm di}}_{\beta,\ell}\left(u(v;q);q\right)
+\frac{\delta_{\alpha,2}}{(k+1)!},
\quad k\geq0.
\eeq
Here the coefficients $b_{\alpha,k}^{\beta,\ell,{\rm di}}$ are given by~\eqref{constb}.
\end{lem}
\begin{prf}
From the expression~\eqref{constb} for $b_{\alpha,k}^{\beta,\ell,\rm di}$ 
we know that the identity~\eqref{thetatheta} can be equivalently written as
\begin{align}
&\theta^{\P}_{1,k}(v(u;q);q)=\frac{1}{k!}
\theta^{X,{\rm di}}_1\left(u;q;k\right)
+\frac{2z}{k!}\left. \frac{\p \theta^{X,{\rm di}}_2\left(u;q;z\right)}{\p z}\right|_{z=k}, \label{221}\\
&\theta^{\P}_{2,k}(v(u;q);q)=\frac{1}{k!}\theta^{X,{\rm di}}_2\left(u;q;k+1\right)+\frac1{(k+1)!}. \label{222}
\end{align}
We know from Lemma~\ref{almostdualityflows} that both sides of~\eqref{222} satisfy the second-order linear PDE~\eqref{feb1}. 
By using the formulae~\eqref{p1theta2} and~\eqref{vu} as well as~\eqref{tloc2}, 
we know that both sides of~\eqref{222} belong to $e^{(k+1)u^1} \CC[[q e^{u^2}]]$.
It then follows from an elementary exercise that the identity~\eqref{222} can be reduced to a finite number of simple identities. 
This proves validity of~\eqref{222}.
In a similar way we can prove~\eqref{221}.
The lemma is proved.
\end{prf}

Let us now prove the following two lemmas.

\begin{lem}\label{lemoo}
For all $k,\ell\geq0$, we have
\beq\label{eqoo}
\Omega^{\P}_{\alpha,k;\beta,\ell}(v; q)
=\sum_{m,n\geq0} b_{\alpha,k}^{\sigma,m,{\rm di}} b_{\beta,\ell}^{\rho,n,{\rm di}}
\Omega^{X,{\rm di}}_{\sigma,m;\rho,n}(u(v;q);q)
+\frac{\eta^{\P}_{\alpha\beta}}{(k+\ell+1)k!\ell!}.
\eeq
\end{lem}
\begin{prf}
Let us first prove~\eqref{eqoo} for $(\alpha,\beta)=(2,2)$. 
From~\eqref{p1theta1}, \eqref{p1theta2}, 
we know that the formula~\eqref{zh-3b} for $\P$ reads
\beq\label{eulerp1}
E\theta^{\P}_{\alpha,k}=(k+\alpha-1)\theta^{\P}_{\alpha,k}+2\delta_{\alpha,1}\theta^{\P}_{2,k-1}.
\eeq
Then by using the formula~\eqref{defomega}, 
one can verify that
\begin{align}
&E\left(
\frac{\eta(\nabla\theta_2^{\P}(v;q;z_1),\nabla\theta_2^{\P}(v;q;z_2))}{z_1+z_2}
\right) \label{Eomega} \\
=&\frac{z_1\p_{z_1}+z_2\p_{z_2}+2}{z_1+z_2}\left(
\eta(\nabla\theta_2^{\P}(v;q;z_1),\nabla\theta_2^{\P}(v;q;z_2))
\right) 
=\sum_{k,\ell\geq0}(k+\ell+2)\Omega^{\P}_{2,k;2,\ell}z_1^k z_2^\ell . 
\end{align}
On the other hand, it follows from~\eqref{thetarecursion} that~\eqref{Eomega}
can also be written as
\begin{align*}
&\frac{E^\sigma}{z_1+z_2}
\left(
\frac{\p^2\theta^{\P}_2(v;q;z_1)}{\p v^\sigma\p v^\alpha}
\eta^{\alpha\beta,\P}
\frac{\p \theta_2^{\P}(v;q;z_2)}{\p v^\beta}
+
\frac{\p\theta^{\P}_2(v;q;z_1)}{\p v^\alpha}
\eta^{\alpha\beta,\P}
\frac{\p^2\theta_2^{\P}(v;q;z_2)}{\p v^\sigma\p v^\beta}
\right)\\
=&\frac{\p \theta_2^{\P}(v;q;z_1)}{\p v^\alpha}
g^{\alpha\beta}(v;q)
\frac{\p \theta_2^{\P}(v;q;z_2)}{\p v^\beta},
\end{align*}
where the coefficients $g^{\alpha\beta}$ are given by the intersection form~\eqref{metric2} for $\P$:
\beq
\left(g^{\alpha\beta}\right)
=
\begin{pmatrix}
	2 q e^{v^2} & v^1 \\
	v^1 & 2
\end{pmatrix}.
\eeq
Hence we obtain that
\beq\label{omega22-1}
\Omega^{\P}_{2,k;2,\ell}(v;q)
= \frac{1}{k+\ell+2}
\frac{\p \theta^{\P}_{2,k}(v;q)}{\p v^\alpha}
g^{\alpha\beta}(v;q)
\frac{\p \theta^{\P}_{2,\ell}(v;q)}{\p v^\beta},
\quad
k,\ell\geq0.
\eeq
It follows from~\eqref{omega22-1} and Lemma~\ref{lemvu} that
\begin{align*}
\Omega_{2,k;2,\ell}^{\P}(v;q)
=&
\left.
\frac{1}{k+\ell+2}
\frac{1}{k!\ell!}
\eta^{\alpha\beta,{\rm di}}
\frac{\p \theta^{X,{\rm di}}_2(u;q;k+1)}{\p u^\alpha}
\frac{\p \theta^{X,{\rm di}}_2(u;q;\ell+1)}{\p u^\beta}
\right|_{u=u(v;q)}\\
=&\sum_{m,n\geq0} b_{2,k}^{\lambda,m,{\rm di}} b_{2,\ell}^{\mu,n,{\rm di}}
\Omega^{X,{\rm di}}_{\lambda,m;\mu,n}(u(v;q);q).
\end{align*}
Here, to derive the last equality we used the definition~\eqref{defomega} for $X$ with diagonal action.
So the identities~\eqref{eqoo} hold true for~$(\alpha,\beta)=(2,2)$.

Next let us prove~\eqref{eqoo} for~$(\alpha,\beta)=(1,2)$.
Similar to the proof of~\eqref{omega22-1}, 
from the formulae~\eqref{thetarecursion}, \eqref{zh-3b} and~\eqref{defomega}
it follows that
\begin{align}
\Omega^{\P}_{1,k;2,\ell}(v;q)
=&\frac{g^{\alpha\beta}}{k+\ell+1}\frac{\p \theta^{\P}_{1,k}(v;q)}{\p v^\alpha}
\frac{\p \theta^{\P}_{2,\ell}(v;q)}{\p v^\beta} 
-\frac{2 g^{\alpha\beta}}{(k+\ell+1)^2}
\frac{\p \theta^{\P}_{2,k-1}(v;q)}{\p v^\alpha}
\frac{\p \theta^{\P}_{2,\ell}(v;q)}{\p v^\beta}.
\end{align}
By using Lemma~\ref{lemvu},
we have
\begin{align}
&\Omega^{\P}_{1,k;2,\ell}(v(u;q);q)
=\frac{\eta^{\alpha\beta,X,\rm di}}{k!\ell!(k+l+1)}
\left(
\frac{\p \theta^{X,{\rm di}}_1(u;q;k)}{\p u^\alpha}
\frac{\p \theta^{X,{\rm di}}_2(u;q;\ell+1)}{\p u^\beta}
\right. 
\nn \\
& \qquad 
\left.+2k\left.\frac{\p^2\theta^{X,{\rm di}}_2(u;q;z)}{\p z \p u^\alpha}\right|_{z=k }
\frac{\p \theta^{X,{\rm di}}_2(u;q;\ell)}{\p u^\beta} 
-\frac{2k }{k+\ell+1}
\frac{\p \theta^{X,{\rm di}}_2(u;q;k)}{\p u^\alpha}
\frac{\p \theta^{X,{\rm di}}_2(u;q;\ell+1)}{\p u^\beta} 
\right)
\nn \\
&\quad   
=\sum_{m,n\geq0} b_{1,k}^{\sigma,m,{\rm di}} b_{2,\ell}^{\rho,n,{\rm di}}
\Omega^{X,{\rm di}}_{\sigma,m;\rho,n}(u;q)
+\frac{1}{(k+\ell+1)k!\ell!},
\end{align}
where 
to derive the last equality we used the definition~\eqref{defomega} for $X$ with diagonal action.
This proves the identities~\eqref{eqoo} for~$(\alpha,\beta)=(1,2)$.

The proof of~\eqref{eqoo} for~$(\alpha,\beta)=(1,1)$ is similar to the above ones,
so we omit it here.
The lemma is proved.
\end{prf}

\begin{lem}\label{lemtopvu}
Let ${\bf t}_{\rm di}({\bf s})$ be defined as in~\eqref{ts}.
Then the following identities
\begin{align}
v_{\rm top}^1({\bf s};q)
=e^{u_{\rm top,di}^{1}({\bf t}_{\rm di}({\bf s});q)}
\left(1+q e^{u_{\rm top,di}^{2}({\bf t}_{\rm di}({\bf s});q)}\right),
\label{topvu}\\
v_{\rm top}^2({\bf s};q)
=2 u_{\rm top,di}^{1}\left({\bf t}_{\rm di}({\bf s});q\right)
+u_{\rm top,di}^{2}\left({\bf t}_{\rm di}({\bf s});q\right),
\label{topvub}
\end{align}
hold true in the ring 
\[V_{\rm di}:=\CC[[s^{1,0}-1,s^{2,0},s^{1,1},s^{2,1},\dots;q]].\]
\end{lem}
\begin{prf}
By using the degree-dimension counting for the GW invariants of $\P$, 
we know that 
\beq
v_{\rm top}^\alpha({\bf s};q)\in\CC[s^{1,0}][[s^{2,0},s^{1,1},s^{2,1},\dots;q]]
\subset
V_{\rm di}.
\eeq
By using the expressions~\eqref{tloc1}, \eqref{tloc2} 
and the definition of the Principal Hierarchy~\eqref{hierarchy}
as well as the formulae~\eqref{311}--\eqref{312},
one can show that the right-hand sides of~\eqref{topvu} and~\eqref{topvub},
denoted by $\tilde v^\alpha({\bf s};q)$,
also belong to $V_{\rm di}$.

Now let us prove the identities~\eqref{topvu}, \eqref{topvub}.
By using the Euler--Lagrange equation~\eqref{eleq}, the formula~\eqref{ts} and Lemma~\ref{lemvu}, 
we have
\begin{align*}
0
=&\sum_{k\geq0}\tilde t_{\rm di}^{\alpha,k}({\bf s}) 
\frac{\p \theta^{X,{\rm di}}_{\alpha,k}}{\p u^\beta}
\left(u_{\rm top,di}({\bf t}_{\rm di}({\bf s});q);q\right)
\\
=&\sum_{k,\ell\geq0}b^{\alpha,k,{\rm di}}_{\sigma,\ell} 
\tilde s^{\sigma,\ell}  
\frac{\p \theta^{X,{\rm di}}_{\alpha,k}}{\p u^\beta}
(u_{\rm top,di}\left({\bf t}_{\rm di}({\bf s});q\right);q)
\\
=& \frac{\p v^\gamma}{\p u^\beta}(u(\tilde v({\bf s};q);q));q)
 \sum_{k\geq0}\tilde s^{\alpha,k} \frac{\p \theta^{\P}_{\alpha,k}}{\p v^\gamma}(\tilde v({\bf s}; q); q),
\end{align*}
where $\left(\frac{\p v^\gamma}{\p u^\beta}\right)$ is the Jacobian of the transformation~\eqref{vu}.
This implies that $\tilde v\left({\bf s};q\right)$
satisfies the genus zero Euler--Lagrange equation~\eqref{eleq} for~$\P$.
Therefore $v_{\rm top}({\bf s};q)=\tilde v({\bf s};q)$.
The lemma is proved.
\end{prf}

From the formula~\eqref{f0omega}, Lemma~\ref{lemoo} 
and Lemma~\ref{lemtopvu},
we arrive at the following proposition.
\begin{prp}
The $g=0$ part of the identity~\eqref{fgfg} holds true.
\end{prp}

\subsection{The anti-diagonal case}

Similarly to the diagonal case, 
we are going to prove the $g=0$ part of the identity~\eqref{fgfg1}.

\begin{lem}\label{lemvu1}
Under the coordinate transformation
\beq\label{vu1}
v^1 = \sqrt{-1} e^{ u^1}(1-2 q e^{u^2}),\quad
v^2 = 2u^1+u^2+\log(1-qe^{u^2}),
\eeq
the following identities hold true:
\beq\label{thetatheta1}
\frac{\p \theta^{\P}_{\alpha,k}(v; q)}{\p v^\beta}
= M_\beta^\gamma(u(v;q);q)
\sum_{\ell\geq0}b_{\alpha,k}^{\sigma,\ell,{\rm ad}}
\frac{\p \theta^{X,\rm ad}_{\sigma,\ell}}{\p u^\gamma}\left(u(v;q);q\right),
\quad k\geq0.
\eeq
Here the coefficients $M_\beta^\gamma=M_\beta^\gamma(u;q)$ are defined by
\beq
\left(M^{\alpha}_\beta\right)_{\alpha,\beta=1,2}=
\begin{pmatrix}
	e^{-u^1} &  \sqrt{-1} q e^{u^2}\\
	- e^{-u^1} & \sqrt{-1} (1-q e^{u^2})
\end{pmatrix},
\eeq
and the coefficients $b_{\alpha,k}^{\beta,\ell,{\rm ad}}$ are given by~\eqref{constb1}.
\end{lem}
\begin{prf}
From the explicit expression~\eqref{constb1} for $b_{\alpha,k}^{\beta,\ell,\rm ad}$ 
we can rewrite~\eqref{thetatheta1} in the form
\begin{align}
&\frac{\p \theta^{\P}_{1,k}}{\p v^\beta}\left(v(u;q);q\right)
=\frac{(\sqrt{-1})^{k-1}}{k!} M_\beta^\gamma \frac{\p}{\p u^\gamma}
\left(\theta^{X,\rm ad}_1(u;q;k)+(2k \p_z+1)\theta^{X,\rm ad}_2(u;q;z)|_{z=k}\right),
\label{thetap1x1}
\\
&\frac{\p \theta^{\P}_{2,k}}{\p v^\beta}\left(v(u;q);q\right)
=\frac{(\sqrt{-1})^{k}}{k!} M_\beta^\gamma \frac{\p \theta^{X,\rm ad}_2(u;q;k+1)}{\p u^\gamma}.
\label{thetap1x2}
\end{align}
For $k=0$, it is easy to see that both sides of~\eqref{thetap1x1} are equal to $\eta_{1\beta}^{\P}$,
and both sides of~\eqref{thetap1x2} are equal to $\eta_{2\beta}^{\P}$.
For $k\geq1$, it follows from the relation~\eqref{eulerp1}
that~\eqref{thetap1x1}, \eqref{thetap1x2} can be further rewritten as
\begin{align}
& \theta^{\P}_{1,k}\left(v(u;q);q\right)
=\frac{(\sqrt{-1})^{k}}{k(k!)}\sum_{\alpha=1}^2\frac{\p}{\p u^\alpha}
\left(\theta^{X,\rm ad}_1(u;q;k)+(2k \p_z-1)\theta^{X,\rm ad}_2(u;q;z)|_{z=k}\right),
\label{ad221}\\
& \theta^{\P}_{2,k}\left(v(u;q);q\right)
=\frac{(\sqrt{-1})^{k+1}}{(k+1)!} \sum_{\alpha=1}^2\frac{\p \theta^{X,\rm ad}_2(u;q;k+1)}{\p u^\alpha}.
\label{ad222}
\end{align}
By using the map~\eqref{vu1}, equation~\eqref{pdeham} and 
Lemma~\ref{almostdualityflows}, we find that 
both sides of~\eqref{ad221} satisfy the following second-order linear 
PDE:
\beq\label{ADpde}
\frac{qe^{u^2}}{1-qe^{u^2}}\frac{\p^2 f(u)}{\p u^1\p u^1}+\frac{\p^2 f(u)}{\p u^2\p u^2}+\frac{qe^{u^2}}{1-qe^{u^2}}\left(\frac{\p f(u)}{\p u^1}-\frac{\p f(u)}{\p u^2}\right)=0.
\eeq
From the formulae~\eqref{p1theta1}, \eqref{vu1} as well as~\eqref{adtloc2},
 we know that both sides of~\eqref{ad222} belong to $e^{(k+1)u^1}\CC[[qe^{u^2}]]$.
It then follows from an elementary exercise that the identity~\eqref{ad222} can be reduced to a finite number of simple identities. 
This proves validity of~\eqref{ad222}.
In a similar way one can prove~\eqref{ad221}.
The lemma is proved. 
\end{prf}

Let us proceed to prove the following two lemmas.

\begin{lem}\label{lemoo1}
For all $k,\ell\geq0$, we have
\beq\label{eqoo1}
\Omega^{\P}_{\alpha,k;\beta,\ell}(v;q)
=-\sum_{m,n\geq0} 
b_{\alpha,k}^{\sigma,m,{\rm ad}} b_{\beta,\ell}^{\rho,n,{\rm ad}}
\Omega^{X,{\rm ad}}_{\sigma,m;\rho,n}(u(v;q);q)
+\eta_{\alpha\beta}^{\P}\frac{(\sqrt{-1})^{k+\ell+1}}{(k+\ell+1)k!\ell!},
\eeq
where $u(v;q)$ is given by~\eqref{vu1}.
\end{lem}
\begin{prf}
The proof is similar to that of Lemma~\ref{lemoo},
so we omit the details.
\end{prf}

\begin{lem}\label{lemtopvu1}
Let ${\bf t}_{\rm ad}\left({\bf s}\right)$ be defined as in~\eqref{ts1}.
The identities
\begin{align}
&v_{\rm top}^1({\bf s};q)
= \sqrt{-1} e^{u_{\rm top,ad}^1\left({\bf t}_{\rm ad}({\bf s});q\right)}
\left(1-2 q e^{u_{\rm top,ad}^2\left({\bf t}_{\rm ad}({\bf s});q\right)}\right),\label{topvu1}\\
&v_{\rm top}^2({\bf s};q)
=2u_{\rm top,ad}^1\left({\bf t}_{\rm ad}({\bf s});q\right)
+u_{\rm top,ad}^2\left({\bf t}_{\rm ad}({\bf s});q\right)
+\log\left(1-q e^{u_{\rm top,ad}^2\left({\bf t}_{\rm ad}({\bf s});q\right)}\right),\label{topvu1b}
\end{align}
hold true in $V_{\rm ad}:=\CC[[s^{1,0}-\sqrt{-1},s^{2,0},s^{1,1},s^{2,1},\dots;q]]$.
\end{lem}
\begin{prf}
Similarly to the proof of Lemma~\ref{lemtopvu},
we can prove that both sides of~\eqref{topvu1}, \eqref{topvu1b} belong to $V_{\rm ad}$. 
By using the genus zero Euler--Lagrange equation~\eqref{eleq}, the formula~\eqref{ts1} and Lemma~\ref{lemvu1},
we obtain that
\begin{align*}
0=&\sum_{k\geq0}\tilde t^{\alpha,k}_{\rm ad}({\bf s}) 
\frac{\p \theta^{X,{\rm ad}}_{\alpha,k}}{\p u^\beta}
(u_{\rm top,ad}({\bf t}_{\rm ad}({\bf s});q);q)\\
=&\sum_{k,\ell\geq0}
b^{\alpha,k,{\rm ad}}_{\sigma,\ell} 
\tilde s^{\sigma,\ell}  
\frac{\p \theta^{X,{\rm ad}}_{\alpha,k}}{\p u^\beta}
(u_{\rm top,ad}({\bf t}_{\rm ad}({\bf s});q);q)\\
 =&-\sqrt{-1}\left(M^{-1}\right)_\gamma^\beta(u(\tilde v({\bf s};q));q)
 \sum_{k\geq0}\tilde s^{\alpha,k}\frac{\p \theta^{\P}_{\alpha,k}}{\p v^\beta}(\tilde v({\bf s};q);q),
\end{align*}
where $\tilde v({\bf s};q)$ denotes the right-hand side of~\eqref{topvu}, \eqref{topvu1}.
Hence $v_{\rm top}({\bf s};q)=\tilde v({\bf s};q)$.
The lemma is proved.
\end{prf}

 From the formula~\eqref{f0omega}, Lemma~\ref{lemoo1} 
and Lemma~\ref{lemtopvu1}, 
we arrive at the following proposition.

 \begin{prp}
 The genus zero part of the identity~\eqref{fgfg1} holds true.
 \end{prp}

\section{Evidence for the genus one and genus two cases}\label{g12}

In this section, we provide evidence for the genus one and genus two parts of the conjectural identities~\eqref{fgfg} and~\eqref{fgfg1}.

\subsection{Review on topological recursion relations in genus one and genus two}

Let us recall the genus one and genus two topological recursion relations given in~\cite{EGX,G2}.
For a smooth algebraic variety $Y$, 
take a homogeneous basis $\phi_1,\dots, \phi_n$  of the cohomology ring of~$Y$.
Denote by $\F_g({\bf t})$ ($g\geq0$) the genus $g$ free energy for the GW invariants of $Y$,
and
\beq
\llangle\tau_{i_1}(\phi_{\alpha_1})\cdots\tau_{i_k}(\phi_{\alpha_k})\rrangle_g
:=\frac{\p^k  \F_g}{\p t^{\alpha_1,i_1}\cdots\p t^{\alpha_k,i_k}},
\quad \mathcal U_\alpha^\beta:=\llangle\phi_\alpha \phi^\beta\rrangle_0,
\eeq
where $\phi^\alpha$ is the dual of $\phi_\alpha$ 
with respect to the Poincar\'e paring.
Define a family of operators 
through the generating series
\beq
\sum_{m\geq0}\D_{\alpha,m}z^m:=\sum_{p\geq0}\left(\left(\frac{\p}{\p t^{1,0}}+z\,\mathcal U\right) \mathcal U\right)_\alpha^\beta \frac{\p}{\p v^{\beta,p}},
\eeq
where $v^{\beta,p}:=\llangle\phi_1^{p+1}\phi^\beta\rrangle_0$
are the jet variables.
It is proved in~\cite{EGX} that
\begin{align}
&\D_{\alpha,k} \F_1
=\left\{
\begin{array}{cc}
\frac1{24}\frac{\p}{\p t^{\alpha,0}}{\rm Tr}\left(\U\right), \quad &k=1,\\
0,\quad  &k>1,
\end{array}
\right. \label{TRR1}\\
&\D_{\alpha,k} \F_2=\R_{\alpha,k}, \quad k\geq2, \label{TRR2}\\
&\left(\D_{\alpha,1}\D_{\beta,1}-3\llangle\phi_\alpha\phi_\beta\phi^\gamma\rrangle_0 \D_{\gamma,1}\right)
\F_2=\R_{\alpha,1;\beta,1}, \label{TRR3}
\end{align}
where $\R_{\alpha,k}$ and $\R_{\alpha,1;\beta,1}$ are given by
\begin{align*}
\R_{\alpha,2}=&
\llangle\phi_\alpha \phi_\beta \phi_\gamma \rrangle_0
\left(\frac7{10}\llangle\phi^\beta\rrangle_1
\llangle \phi^\gamma \rrangle_1
+\frac1{10}\llangle\phi^\beta\phi^\gamma\rrangle_1\right)
+\frac{13}{240}\llangle\phi_\alpha \phi_\beta \phi_\gamma \phi^\gamma\rrangle_0 
\llangle\phi^\beta\rrangle_1
\\
&-\frac1{240}\llangle\phi_\alpha\phi^\beta\rrangle_1
\llangle\phi_\beta\phi_\gamma\phi^\gamma\rrangle_0
+\frac1{960}\llangle\phi_\alpha\phi_\beta\phi^\beta\phi_\gamma\phi^\gamma\rrangle_0,\\
\R_{\alpha,3}=&
\llangle\phi_\alpha\phi_\beta\phi_\gamma\rrangle_0
\left(
\frac1{20}\llangle\phi^\beta\rrangle_1
\llangle\phi^\gamma\phi^\sigma\phi_\sigma\rrangle_0
+\frac1{480}\llangle\phi^\beta\phi^\gamma\phi^\sigma\phi_\sigma\rrangle_0
\right)\\
&+\frac1{1152}\llangle\phi_\alpha\phi^\beta\phi^\gamma\phi_\gamma\rrangle_0
\llangle\phi_\beta\phi^\sigma\phi_\sigma\rrangle_0,\\
\R_{\alpha,4}=&
\frac1{1152}\llangle\phi_\alpha\phi^\beta\phi^\gamma\rrangle_0
\llangle\phi_\beta\phi_\gamma\phi^\sigma\rrangle_0
\llangle\phi_\sigma\phi^\delta\phi_\delta\rrangle_0,
\quad \R_{\alpha,k}=0\,(k>4),\\
\R_{\alpha,1;\beta,1}
=&\frac{13}{10}
\llangle\phi_\alpha\phi_\beta\phi_\gamma\phi_\sigma\rrangle_0
\llangle\phi^\gamma\rrangle_1
\llangle\phi^\sigma\rrangle_1
+\frac45
\left(
\llangle\phi_\alpha\phi_\gamma\rrangle_1
\llangle\phi_\sigma\rrangle_1
+\frac1{24}
\llangle\phi_\alpha\phi_\gamma\phi^\sigma\rrangle_1
\right)
\llangle\phi_\beta\phi^\gamma\phi^\sigma\rrangle_0\\
&+\frac45
\llangle\phi_\alpha\phi^\gamma\phi^\sigma\rrangle_0
\left(
\llangle\phi_\beta\phi_\gamma\rrangle_1
\llangle\phi_\sigma\rrangle_1
+\frac1{24}
\llangle\phi_\beta\phi_\gamma\phi_\sigma\rrangle_1
\right)+\frac1{48}
\llangle\phi_\alpha\phi_\gamma\phi_\sigma\phi^\sigma\rrangle_0
\llangle\phi^\gamma\phi_\beta\rrangle_1
\\
&-\frac45
\llangle\phi_\alpha\phi_\beta\phi_\gamma\rrangle_0
\left(
\llangle\phi^\gamma\phi_\sigma\rrangle_1
\llangle\phi^\sigma\rrangle_1
+\frac1{24}
\llangle\phi^\gamma\phi_\sigma\phi^\sigma\rrangle_1
\right)+\frac1{48}
\llangle\phi_\alpha\phi_\gamma\rrangle_1
\llangle\phi_\beta\phi^\gamma\phi_\sigma\phi^\sigma\rrangle_0\\
&+\frac{23}{240}
\llangle\phi_\alpha\phi_\beta\phi_\gamma\phi_\sigma\phi^\sigma\rrangle_0
\llangle\phi^\gamma\rrangle_1
-\frac1{80}
\llangle\phi_\alpha\phi_\beta\phi_\gamma\rrangle_1
\llangle\phi^\gamma\phi^\sigma\phi_\sigma\rrangle_0\\
&+\frac7{30}
\llangle\phi_\alpha\phi_\beta\phi_\gamma\phi_\sigma\rrangle_0
\llangle\phi^\gamma\phi^\sigma\rrangle_1
+\frac1{576}
\llangle\phi_\alpha\phi_\beta\phi_\gamma\phi^\gamma\phi_\sigma\phi^\sigma\rrangle_0.
\end{align*}

We are going to show that
for the resolved conifold $X$ with diagonal and anti-diagonal actions,
the genus one and genus two free energies obtained from Conjectures~\ref{cnjmain}, \ref{cnjmain1}
satisfy the above topological recursion relations~\eqref{TRR1}--\eqref{TRR3}.
This will give evidences for the validity of these two conjectures.
We know from~\cite{DW,DZ-norm,DZ1,EYY} the following lemma.

\begin{lem}\label{jetrep}
For $g\geq1$, 
there exist functions $F_g^{\P}(v_0,v_1,\dots,v_{3g-2};q)$,
where $v_k=(v^{1}_k, v^{2}_k)$ are pairs of indeterminates,
such that 
\beq\label{p1jetrep}
\fcp_g({\bf s};q)=F_g^{\P}\left(v_{\rm top}({\bf s};q),\frac{\p v_{\rm top}({\bf s};q)}{\p s^{1,0}},\dots,\frac{\p^{3g-2} v_{\rm top}({\bf s};q)}{\p (s^{1,0})^{3g-2}};q\right).
\eeq
Similarly,
for $g\geq1$, 
there exist functions $F_g^{X,{\rm di/ad}}(u_0,u_1,\dots;q)$, 
where $u_k=(u_k^1, u_k^2)$ are pairs of indeterminates, 
such that  
\beq
\F^{X, \rm di/ad}_g({\bf t};q)
=F_g^{X, \rm di/ad}\left(u_{\rm top,di/ad}({\bf t};q),\frac{\p u_{\rm top,di/ad}({\bf t};q)}{\p t^{1,0}},\dots;q\right).
\eeq
\end{lem}

\subsection{Evidence for the diagonal case}

Let us now consider the diagonal case. 
It follows from Lemma \ref{jetrep} that both sides of~\eqref{fgfg} 
admit jet variable representations. As we mentioned in the Introduction, unlike the non-invertibility of the transformation~\eqref{ts}, 
the jet variables between the two models are invertible. 
To be precise, denote 
\beq
v^\alpha=v^\alpha_0:=v^\alpha_{\rm top},
\
u^\alpha=u^\alpha_0:=u^\alpha_{\rm top,di},
\
v^{\alpha}_k:=\frac{\p^k v_{\rm top}^\alpha}{\p (s^{1,0})^k},
\
u^{\alpha}_k:=\frac{\p^k u_{\rm top,\rm di}^\alpha}{\p (t^{1,0})^k},
\
k\geq1,
\eeq
then Lemma~\ref{lemtopvu} reads
\begin{align}
&v^1=  e^{u^1}\left(1+q e^{u^2}\right),\quad
v^2= 2u^1+u^2. \label{vu0}
\end{align}
By applying $\frac{\p}{\p s^{1,0}}$ to both sides of the above equations and 
by using~\eqref{ts},
we obtain that
\beq
v^1_1=e^{u^1}\left((1+q e^{u^2})u^1_1+ q e^{u^2}u^2_1\right),\quad
v^2_1=2 u^1_1+ u^2_1.
\eeq
In general, we have the recursion relation
\beq\label{vuk}
v^\alpha_k=\sum_{\ell\geq1}u^\beta_\ell \frac{\p v^{\alpha}_{k-1}}{\p u^\beta_{\ell-1}},
\quad k\geq1.
\eeq 

It follows from~\eqref{p1jetrep} that for $g\geq1$, the left-hand side of~\eqref{fgfg} can be represented in terms of the jet variables $v^\alpha_k$.
Then substituting~\eqref{vu0} and \eqref{vuk} into this representation,
we obtain the predicted expressions of $\F_g^{X,\rm di}$ $(g\geq1)$ in terms of jet variables~$u^\alpha_k$.
For example, 
the genus one free energy $\F_1^{X,\rm di}$ obtained from~\eqref{fgfg} has the following expression:
\beq\label{expF1di}
\F_1^{X,{\rm di}}
=\frac1{24}\log D_{\rm di}
-\frac1{12}\li_1(qe^{u^2})-\frac1{24}u^2.
\eeq
Here
\beq
D_{\rm di}:=(u^1_1)^2-2\xi u^1_1 u^2_1-\xi (u^2_1)^2,\quad
\xi:=\frac{qe^{u^2}}{1-qe^{u^2}},
\eeq
and we used the explicit expression for~$\F^{\P}_1$~\cite{DZ2,DZ-norm}:
\beq\label{expf1}
\F^{\P}_1=
\frac1{24}\log\left( \left(v^1_1\right)^2 - q e^{v^2} \left(v^2_1\right)^2\right)-\frac{v^2}{24}.
\eeq
Similarly, from the explicit expression for $\F^{\P}_2$ (cf.~\cite{DY1,DZ2,DZ-norm}) and the $g=2$ case of~\eqref{fgfg},
we arrive at
\begin{align}
F^{X,\rm di}_2=
&\frac{\xi ^4 (\xi +1)^4 \left(64 \xi ^3+80 \xi ^2+24 \xi +1\right) (u^2_1)^{10}}{90 D_{\rm di}^4} \nn\\
&+\frac{4  \xi ^4 (\xi +1)^4 \left(16 \xi ^3+24 \xi ^2+10 \xi +1\right) u^1_1 (u^2_1)^9}{45 D_{\rm di}^4} \nn \\
&+\frac{ \xi ^3 (\xi +1)^3 \left(4096 \xi ^4+13888 \xi ^3+12240 \xi ^2+3160 \xi +121\right) (u^2_1)^8}{5760 D_{\rm di}^3}+\cdots.
\label{expF2di}
\end{align} 
Here we omit the explicit expressions of the remaining 98 terms.

By using the explicit expressions~\eqref{expF1di}, \eqref{expF2di} of $\F_1^{X,\rm di}$ and $\F_2^{X,\rm di}$, we arrive at the following proposition.

\begin{prp}
The genus one and genus two free energies 
$\F^{X,{\rm di}}_{1}$ and $\F^{X,{\rm di}}_{2}$ 
obtained from~\eqref{fgfg}
satisfy the topological recursion relations~\eqref{TRR1}--\eqref{TRR3}.
\end{prp}

By using these expressions,
we obtain in particular
the primary genus one and genus two free energies:
\begin{align}
& \F_1^{X, \rm di}\big|_{t^{\alpha,k}=t^\alpha\delta^{k,0}} 
=-\frac1{24}t^{2}-\frac1{12}{\rm Li}_1\left(qe^{t^2}\right),\\
& \F_2^{X, \rm di}\big|_{t^{\alpha,k}=t^\alpha\delta^{k,0}}  
=-\frac{1-14qe^{t^{2}}+q^2 e^{2t^{2}}}{2880\left(1-qe^{t^{2}}\right)^2}
=-\frac{1}{2880}+\frac{1}{240} {\rm Li}_{-1}\left(qe^{t^{2}}\right).
\end{align}

\subsection{Evidence for the anti-diagonal case}

Let us consider the anti-diagonal case.
Denote
\beq
v^\alpha=v^\alpha_0:=v^\alpha_{\rm top},
\
u^\alpha=u^\alpha_0:=u^\alpha_{\rm top,ad},
\
v^{\alpha}_k:=\frac{\p^k v_{\rm top}^\alpha}{\p (s^{1,0})^k},
\
u^{\alpha}_k:=\frac{\p^k u_{\rm top,\rm ad}^\alpha}{\p (t^{1,0})^k},
\
k\geq1,
\eeq
then Lemma~\ref{lemtopvu1} reads
\begin{align}
&v^1=  \sqrt{-1}e^{u^1}\left(1-2q e^{u^2}\right),\quad
v^2= 2u^1+u^2+\log\left(1-qe^{u^2}\right). \label{vu0a}
\end{align}
By applying $\frac{\p}{\p s^{1,0}}$ to both sides of the above equations and 
by using~\eqref{ts1},
we obtain
\beq
v^1_1=e^{ u^1}(1-4qe^{u^2})u^1_1- e^{ u^1+u^2}
\frac{3-4qe^{u^2}}{1-qe^{u^2}}u^2_1,\quad
v^2_1=\frac{3-4qe^{u^2}}{1-qe^{u^2}}u^1_1+\frac{1-4qe^{u^2}}{1-qe^{u^2}}u^2_1.
\eeq
In general, we have the recursion relation
\beq\label{vuka}
v^\alpha_k=\sum_{\ell\geq1}u^\beta_\ell \frac{\p v^\alpha_{k-1}}{\p u^\beta_{\ell-1}} 
+\sum_{m\geq0}\frac{\p v^\alpha_{k-1}}{\p u^\gamma_m}
\left(\sum_{\ell\geq1}u^\beta_\ell \frac{\p }{\p u^\beta_{\ell-1}}\right)^m\left(-\frac{q e^{u^2} u^2_1}{1-q e^{u^2}}\delta^{\gamma,1}+u^1_1\delta^{\gamma,2}\right)
,\quad k\geq1.
\eeq 

By substituting~\eqref{vu0a} and \eqref{vuka} into the jet representation of the left-hand side of~\eqref{fgfg1},
we obtain the predicted expressions of $\F_g^{X,\rm ad}$ $(g\geq1)$ in terms of jet variables~$u^\alpha_k$.
For example,
the genus one and genus two free energies $\F_{1}^{X,\rm ad}$, $\F_{2}^{X,\rm ad}$ obtained from~\eqref{fgfg1} 
have the following expressions:
\begin{align}
\F_1^{X, \rm ad}=&\frac1{24}\log D_{\rm ad}
-\frac1{12}\log(1-qe^{u^2})-\frac1{24}u^2, \label{expF1ad} \\
\F_2^{X, \rm ad}=
&-\frac{\xi^4  (\xi+1)^3\left(u^2_1\right)^{10}}{90 D_{\rm ad}^4} 
-\frac{ \xi^4  (\xi+1)^2 \left(u^2_1\right)^8 u^2_2}{15 D_{\rm ad}^4}
+\cdots,\label{expF2ad}
\end{align}
where
\beq
D_{\rm ad}:=\left(u^1_1\right)^2+\xi  \left(u^2_1\right)^2,
\quad
\xi:=\frac{qe^{u^2}}{1-qe^{u^2}},
\eeq
and we omit the explicit expressions of the remaining 51 terms in~\eqref{expF2ad}.

By using the explicit expressions~\eqref{expF1ad}, \eqref{expF2ad} of $\F_1^{X,\rm ad}$ and $\F_2^{X,\rm ad}$, we arrive at the following proposition.

\begin{prp}
The genus one and genus two free energies 
$\F^{X,{\rm ad}}_{1}$ and $\F^{X,{\rm ad}}_{2}$ 
obtained from~\eqref{fgfg1}
satisfy the topological recursion relations~\eqref{TRR1}--\eqref{TRR3}.
\end{prp}

In particular, by taking $t^{\alpha,k\geq1}=0$,
we obtain the primary parts of $\F_1^{X,\rm ad}$ and $\F_2^{X,\rm ad}$ 
\begin{align*}
&\F_1^{X,{\rm ad}}|_{t^{\alpha,k}=t^\alpha\delta^{k,0}}
=-\frac1{24}t^2+\frac1{12}{\rm Li}_1 \bigl(qe^{t^2}\bigr),
\\
&\F_2^{X,{\rm ad}}|_{t^{\alpha,k}=t^\alpha\delta^{k,0}}
=\frac{1+10 q e^{t^{2}}+q^2 e^{2t^{2}}}{2880\left(1-q e^{t^{2}}\right)^2}
=\frac1{2880}+\frac1{240}{\rm Li}_{-1} \bigl(qe^{t^2}\bigr),
\end{align*}
which agree with the ones given in~\cite{Bri}:
\beq
\F_g^{X,\rm ad}|_{t^{\alpha,k}=t^\alpha\delta^{k,0}} =\frac{|B_{2g}|}{2g(2g-2)!}{\rm Li}_{3-2g}\left(qe^{t^2}\right)+\frac{|B_{2g}B_{2g-2}|}{2g(2g-2)(2g-2)!},
\quad
g\geq2.
\eeq

\section{Further remarks}\label{section5}

We know that the genus one free energy can be represented in the following form (cf.~\cite{DW, DZ3, G}):
\beq
\F_1=\left.\left(\frac1{24}\log {\rm det}\left(c^\alpha_{\beta\gamma}v_x^\gamma\right)+G_1(v)\right)\right|_{v=v_{\rm top}},
\eeq
and the genus two free energy can be represented in the form~\cite{DLZ}:
\beq
\F_2=\left.\left(\sum_{i=1}^{16}c_i Q_i+G_2(v,v_x,v_{xx})\right)\right|_{v=v_{\rm top}},
\eeq
where $c_i$ are certain constants, 
$Q_i$ are given by some genus zero and genus one correlation functions corresponding to the dual graphs of some stable algebraic curves,
and $G_1(v)$, $G_2(v,v_x,v_{xx})$ are called the genus one and genus two $G$-functions respectively.

Our first remark is that
from the explicit expressions~\eqref{expF1di}, \eqref{expF2di}, \eqref{expF1ad} and~\eqref{expF2ad}, 
it follows that the corresponding genus one and genus two $G$-functions have the expressions
\begin{align*}
G_1^{X,\rm di}=&\frac1{12}\log(1-qe^{u^2})-\frac1{24}u^2,\\
G_2^{X,\rm di}
=&\frac{12\xi^2+12\xi-1}{2880}D_{\rm di}
+\frac{2\xi(1+\xi)(1+2\xi)}{2880}u^1_1 u^2_1 \\
&
-\frac{\xi(1+\xi)(1+62\xi+64\xi^2)}{2880}\left(u^2_1\right)^2
+\frac{\xi(1+\xi)}{80}u^1_{2}
-\frac{\xi(1+\xi)(4+17\xi)}{720}u^2_{2}
\\&
-\frac{\xi^2 (1+\xi)^2(1+2\xi)}{180}\frac{u^1_1\left(u^2_1\right)^3}{D_{\rm di}}
+\frac{\xi(1+\xi)(1+2\xi)}{320}\frac{u^1_1 u^1_{2} u^2_1}{D_{\rm di}}
\\&
-\frac{\xi(1+\xi)(1+24\xi+32\xi^2)}{2880}\frac{u^1_1 u^2_1 u^2_{2}}{D_{\rm di}}
-\frac{ \xi^2(1+\xi)(3+11\xi+8\xi^2)}{1440}\frac{\left(u^2_1\right)^4}{D_{\rm di}}
\\&
+\frac{\xi(1+\xi)(1+8\xi)}{2880}\frac{u^1_{2} \left(u^2_1\right)^2 }{D_{\rm di}}
-\frac{ \xi^2(1+\xi)(1+2\xi)}{320}\frac{\left(u^2_1\right)^2 u^2_{2}}{D_{\rm di}},
\\
G_1^{X,\rm ad}=&-\frac1{12}\log(1-qe^{u^2})-\frac1{24}u^2,
\\
G_2^{X,\rm ad}=&
\frac{\xi (\xi+1) u^2_1}{1440D_{\rm ad}} \left(2  \xi (\xi+1) \left(u^2_1\right)^3+3  \xi u^2_1 u^2_{2}+3 u^1_1 u^1_{2}\right)
\\&
-\frac{ \xi}{2880} \left(16 (\xi+1) u^2_{2}+\left(26 \xi^2+25 \xi+1\right) \left(u^2_1\right)^2\right)
+\frac{ 12 \xi^2+12 \xi+1}{2880}D_{\rm ad}.
\end{align*}

Our second remark is that
the genus one and genus two free energies 
$\F_{1,2}^{X,\rm di}$ and $\F_{1,2}^{X,\rm ad}$ 
obtained from~\eqref{fgfg} and~\eqref{fgfg1}
also satisfy the following Belorousski--Pandharipande equations~\cite{BP} up to $0\leq k_1,k_2,k_3\leq 1$:
\begin{align*}
%1
-&2\llangle\phi_\alpha\rrangle_2
\llangle\phi^\alpha\phi_\beta\tau_{k_1}(\phi_{\alpha_1})\rrangle_0
\llangle\phi^\beta\tau_{k_2}(\phi_{\alpha_2})\tau_{k_3}(\phi_{\alpha_3})\rrangle_0\\
%2
+&2\left(\llangle\tau_1(\phi_{\alpha})\rrangle_2
\llangle\phi^\alpha\tau_{k_1}(\phi_{\alpha_1})\tau_{k_2}(\phi_{\alpha_2})\tau_{k_3}(\phi_{\alpha_3})\rrangle_0
-\llangle\phi_\alpha\rrangle_2 \llangle\phi^\alpha\phi_\beta\rrangle_0
 \llangle\phi^\beta\tau_{k_1}(\phi_{\alpha_1})\tau_{k_2}(\phi_{\alpha_2})\tau_{k_3}(\phi_{\alpha_3})\rrangle_0\right)\\
%3
+&3\left(\llangle\phi_\alpha \tau_{\alpha_1,k_1+1}\rrangle_2 
\llangle\phi^\alpha\tau_{k_2}(\phi_{\alpha_2})\tau_{k_3}(\phi_{\alpha_3})\rrangle_0
-\llangle\phi_\alpha\tau_{k_1}(\phi_{\alpha_1})\rrangle_0
\llangle\phi^\alpha\phi_\beta\rrangle_2
\llangle\phi^\beta\tau_{k_2}(\phi_{\alpha_2})\tau_{k_3}(\phi_{\alpha_3})\rrangle_0\right)\\
%4
-&3\left(\llangle\tau_{\alpha,1}\tau_{k_1}(\phi_{\alpha_1})\rrangle_2
\llangle\phi^\alpha\tau_{k_2}(\phi_{\alpha_2})\tau_{k_3}(\phi_{\alpha_3})\rrangle_0
-\llangle\phi_\alpha\tau_{k_1}(\phi_{\alpha_1})\rrangle_2
 \llangle\phi^\alpha\phi_\beta\rrangle_0
 \llangle\phi^\beta\tau_{k_2}(\phi_{\alpha_2})\tau_{k_3}(\phi_{\alpha_3})\rrangle_0 \right)\\
%5
+&\frac15\llangle\phi_\alpha\rrangle_1 
\llangle\phi^\alpha\phi_\beta\tau_{k_1}(\phi_{\alpha_1})\tau_{k_2}(\phi_{\alpha_2})\tau_{k_3}(\phi_{\alpha_3})\rrangle_0
 \llangle\phi^\beta\rrangle_1
%6
-\frac65\llangle\phi_\alpha\rrangle_1
 \llangle\phi^\alpha \phi_\beta \tau_{k_1}(\phi_{\alpha_1})\tau_{k_2}(\phi_{\alpha_2}) \rrangle_0
  \llangle\phi^\beta\tau_{k_3}(\phi_{\alpha_3})\rrangle_1 \\
%7
+&\frac{12}5\llangle\phi_\alpha\rrangle_1
\llangle\phi^\alpha \phi_\beta \tau_{k_1}(\phi_{\alpha_1}) \rrangle_0 
\llangle\phi^\beta\tau_{k_2}(\phi_{\alpha_2})\tau_{k_3}(\phi_{\alpha_3})\rrangle_1
%8
-\frac{18}5\llangle\phi_\alpha\tau_{k_1}(\phi_{\alpha_1})\rrangle_1
 \llangle \phi^\alpha\phi_\beta\tau_{k_2}(\phi_{\alpha_2})\rrangle_0
 \llangle\phi^\beta \tau_{k_3}(\phi_{\alpha_3})\rrangle_1\\
%9
-&\frac65\llangle\phi_\alpha\rrangle_1
\llangle\phi^\alpha\phi_\beta\rrangle_1
\llangle\phi^\beta\tau_{k_1}(\phi_{\alpha_1})\tau_{k_2}(\phi_{\alpha_2})\tau_{k_3}(\phi_{\alpha_3})\rrangle_0
%10
+\frac95\llangle\phi_\alpha\tau_{k_1}(\phi_{\alpha_1})\rrangle_1
\llangle\phi^\alpha\phi_\beta\rrangle_1
\llangle\phi^\beta\tau_{k_2}(\phi_{\alpha_2})\tau_{k_3}(\phi_{\alpha_3})\rrangle_0\\
%11
-&\frac65\llangle\phi_\alpha\rrangle_1
\llangle\phi^\alpha\phi_\beta \tau_{k_1}(\phi_{\alpha_1}) \rrangle_1 \llangle\phi^\beta\tau_{k_2}(\phi_{\alpha_2})\tau_{k_3}(\phi_{\alpha_3})\rrangle_0
%12
+\frac1{120}\llangle\phi_\alpha \phi^\alpha\phi_\beta \tau_{k_1}(\phi_{\alpha_1})\tau_{k_2}(\phi_{\alpha_2})\tau_{k_3}(\phi_{\alpha_3})\rrangle_0
 \llangle\phi^\beta\rrangle_1\\
%13
-&\frac3{40}\llangle\phi_\alpha \phi^\alpha\phi_\beta \tau_{k_1}(\phi_{\alpha_1})\tau_{k_2}(\phi_{\alpha_2})\rrangle_0
 \llangle\phi^\beta\tau_{k_3}(\phi_{\alpha_3})\rrangle_1
%14
+\frac3{40}\llangle\phi_\alpha \phi^\alpha\phi_\beta \tau_{k_1}(\phi_{\alpha_1})\rrangle_0
 \llangle\phi^\beta\tau_{k_2}(\phi_{\alpha_2})\tau_{k_3}(\phi_{\alpha_3})\rrangle_1\\
%15
-&\frac1{120}\llangle\phi_\alpha \phi^\alpha\phi_\beta\rrangle_0
 \llangle\phi^\beta \tau_{k_1}(\phi_{\alpha_1}) \tau_{k_2}(\phi_{\alpha_2})\tau_{k_3}(\phi_{\alpha_3})\rrangle_1
%16
+\frac1{10}\llangle\phi_\alpha\phi_\beta  \tau_{k_1}(\phi_{\alpha_1}) \tau_{k_2}(\phi_{\alpha_2})\tau_{k_3}(\phi_{\alpha_3})\rrangle_0
 \llangle\phi^\alpha\phi^\beta\rrangle_1\\
%17
-&\frac3{10}\llangle\phi_\alpha\phi_\beta  \tau_{k_1}(\phi_{\alpha_1}) \tau_{k_2}(\phi_{\alpha_2})\rrangle_0
 \llangle\phi^\alpha\phi^\beta\tau_{k_3}(\phi_{\alpha_3})\rrangle_1
%18
+\frac1{10}\llangle\phi_\alpha\phi_\beta  \tau_{k_1}(\phi_{\alpha_1}) \rrangle_0
 \llangle\phi^\alpha\phi^\beta\tau_{k_2}(\phi_{\alpha_2})\tau_{k_3}(\phi_{\alpha_3})\rrangle_1 \nn \\
%19
-&\frac1{20}\llangle\phi_\alpha\phi^\alpha\phi_\beta \rrangle_1
 \llangle\phi^\beta\tau_{k_1}(\phi_{\alpha_1})\tau_{k_2}(\phi_{\alpha_2})\tau_{k_3}(\phi_{\alpha_3})\rrangle_0
%20
-\frac1{20}\llangle\phi_\alpha\phi^\alpha\phi_\beta\tau_{k_1}(\phi_{\alpha_1}) \rrangle_1
 \llangle\phi^\beta\tau_{k_2}(\phi_{\alpha_2})\tau_{k_3}(\phi_{\alpha_3})\rrangle_0 \nn \\
+&(1\leftrightarrow2\leftrightarrow3)=0. \label{bpeq}
\end{align*}
Here
$(1\leftrightarrow2\leftrightarrow3)$ denotes the other terms obtained by permutating $(\alpha_1,k_1),(\alpha_2,k_2),(\alpha_3,k_3)$.

\begin{appendices}

\section{Some related calculations}

The first several coefficients of the deformed coordinates are given in Table~\ref{tab1}.
\begin{table}[ht]\label{tab1}
\centering
\begin{tabular}{|l|c|c|c|}\hline
$(\alpha,k)$ &  $\theta^{\P}_{\alpha,k}$ & $\theta^{X,{\rm di}}_{\alpha,k}$ & $\theta^{X,{\rm ad}}_{\alpha,k}$ \\ \hline
$(1,0)$ & $v^2$ & $2u^1 +u^2$ & $u^2$  \\ \hline
$(2,0)$ & $v^1$ & $u^1$ & $u^1$ \\ \hline
$(1,1)$ & $v^1 v^2$ & 
$(u^1)^2 +u^1 u^2$ & $u^1 u^2$\\\hline
$(2,1)$ & $\frac{\left(v^1\right)^2}2+q e^{v^2}$ &
$\frac{(u^1)^2}{2}+ {\rm Li}_2\big(qe^{u^2}\big)$ &
$\frac{(u^1)^2}{2}- {\rm Li}_2\big(qe^{u^2}\big)$ \\\hline
$(1,2)$ & $\frac{\left(v^1\right)^2 v^2}2+q e^{v^2}\left(v^2-2\right)$ &
\tabincell{c}{$\frac1{3} (u^1)^3+\frac12(u^1)^2 u^2
+ u^2{\rm Li}_2\big(qe^{u^2}\big)$\\
$-2 {\rm Li}_3\big(qe^{u^2}\big)$} &
\tabincell{c}{$\frac12(u^1)^2 u^2
- u^2{\rm Li}_2\big(qe^{u^2}\big)$\\
$+2 {\rm Li}_3\big(qe^{u^2}\big)$}\\\hline
$(2,2)$ & $\frac{\left(v^1\right)^3}6+q v^1 e^{v^2}$ & 
\tabincell{c}{
$\frac16(u^1)^3+  u^1{\rm Li}_2(qe^{u^2})$
}
&
\tabincell{c}{
$\frac16(u^1)^3-u^1{\rm Li}_2(qe^{u^2})$
}
\\\hline
\end{tabular}
\caption{First several $\theta_{\alpha,k}$}
\end{table}

We also list some two-point correlations functions in genus zero as follows:
\begin{align*}
&\Omega_{1,1;1,1}^{\P}
=(v^1)^2v^2+2qe^{v^2}-2qv^2e^{v^2}+q(v^2)^2e^{v^2},\\
&\Omega_{1,1;2,1}^{\P}
=\frac13(v^1)^3+qv^1v^2e^{v^2},\\
&\Omega_{2,1;2,1}^{\P}
=q(v^1)^2e^{v^2}+\frac12q^2e^{2v^2},\\
&\Omega_{1,1;1,1}^{X,\rm di}
=\frac23(u^1)^3+(u^1)^2u^2+2\li_3(qe^{u^2})-2u^2\li_2(qe^{u^2})+(u^2)^2\li_1(qe^{u^2}),\\
&\Omega_{1,1;2,1}^{X,\rm di}
=\frac13(u^1)^3+u^1u^2\li_1(qe^{u^2})-(\li_1(qe^{u^2}))^2,\\
&\Omega_{2,1;2,1}^{X,\rm di}
=\left((u^1)^2-2u^1\li_1(qe^{u^2})\right)\li_1(qe^{u^2})- (\li_1(qe^{u^2}))^3
+\int_{-\infty}^{u^2}\frac{(\li_1(qe^y))^2dy}{1-qe^y},\\
&\Omega_{1,1;1,1}^{X,\rm ad}
=(u^1)^2u^2-2\li_3(qe^{u^2})+2u^2\li_2(qe^{u^2})-(u^2)^2\li_1(qe^{u^2}),\\
&\Omega_{1,1;2,1}^{X,\rm ad}
=\frac13(u^1)^3-u^1u^2\li_1(qe^{u^2}),\\
&\Omega_{2,1;2,1}^{X,\rm ad}
=-(u^1)^2\li_1(qe^{u^2})+\int_{-\infty}^{u^2}(\li_1(qe^y))^2dy.
\end{align*}

By using the genus zero Euler--Lagrange equation~\eqref{eleq}
and the data in Table~\ref{tab1},
we obtain
the first several terms in the topological solution $v_{\rm top}$ as follows:
\begin{align*}
v^1_{\rm top}=
& s^{1,0}+q s^{2,1}+s^{1,0} s^{1,1}+q s^{1,1} s^{2,1}+q s^{2,0}s^{2,1} 
+s^{1,0}\left(s^{1,1}\right)^2 + q\left(s^{1,1}\right)^2 s^{2,1} \\
& +2q s^{1,1} s^{2,0} s^{2,1}+\frac q2\left(s^{2,0}\right)^2s^{2,1}
+q s^{1,0}\left(s^{2,1}\right)^2+q^2 s^{2,0} \left(s^{2,1}\right)^3
+ \cdots, \\
v^2_{\rm top}=
&s^{2,0}+ s^{1,1}s^{2,0}+s^{1,0}s^{2,1}+q \left(s^{2,1}\right)^2
+\left(s^{1,1}\right)^2s^{2,0}
+2 s^{1,0} s^{1,1} s^{2,1}\\
&+2 q s^{1,1}\left(s^{2,1}\right)^2
+q s^{2,0} \left(s^{2,1}\right)^2+\cdots.
\end{align*}
Similarly, the topological solutions $u_{\rm top,di}$ and~$u_{\rm top,ad}$ have the form
\begin{align*}
u^1_{\rm top,di}
=&t^{1,0}
- \log(1-q) t^{2,1}
+t^{1,0}t^{1,1}
-\log(1-q)t^{1,1}t^{2,1}
+\frac{ q}{1-q}t^{2,0} t^{2,1}\\
&+\frac{2\kappa^3 q\log(1-q)}{1-q}\left(t^{2,1}\right)^2
+t^{1,0}\left(t^{1,1}\right)^2
- \log(1-q) \left(t^{1,1}\right)^2 t^{2,1}
+\frac{2 q}{1-q}t^{1,1}t^{2,0}t^{2,1}\\
&+\frac{ q}{2(1-q)^2} \left(t^{2,0}\right)^2 t^{2,1}
+\frac{4 q \log(1-q)}{1-q} t^{1,1} \left(t^{2,1}\right)^2
+\frac{ q}{1-q}t^{1,0}\left(t^{2,1}\right)^2\\
&+\frac{2 q (1-q) \log(1-q)}{(1-q)^2}t^{2,0}\left(t^{2,1}\right)^2
+\frac{ q\log(1-q)(2\log(1-q)-1-3q)}{(1-q)^2}\left(t^{2,1}\right)^3
+\cdots, \\
u^2_{\rm top,di}
=&t^{2,0}
+2\log(1-q) t^{2,1}
+t^{1,1}t^{2,0}
+t^{1,0}t^{2,1}
+2\log(1-q)t^{1,1}t^{2,1}\\
&-\frac{2 q}{1-q}t^{2,0}t^{2,1}
-\frac{(1+3q)\log(1-q)}{1-1}\left(t^{2,1}\right)^2
+\left(t^{1,1}\right)^2 t^{2,0}
+2t^{1,0}t^{1,1}t^{2,1}\\
&+2\log(1-q) \left(t^{1,1}\right)^2 t^{2,1}
-\frac{4 q}{1-q} t^{1,1}t^{2,0}t^{2,1}
-\frac{ q}{(1-q)^2}\left(t^{2,0}\right)^2 t^{2,1}
-\frac{2 q}{1-q} t^{1,0}\left(t^{2,1}\right)^2\\
&-\frac{2(1+3q)\log(1-q)}{1-q}t^{1,1}\left(t^{2,1}\right)^2
+\frac{\left(1+3q-4\log(1-q)\right)}{(1-q)^2}t^{2,0} \left(t^{2,1}\right)^2\\
&+\frac{4 q \left(1+q-\log(1-q)\right)\log(1-q)}{(1-q)^2}\left(t^{2,1}\right)^3
+\cdots,\\
u^1_{\rm top,ad}
=&t^{1,0}
+\log(1-q) t^{2,1}
+t^{1,0}t^{1,1}
+\log(1-q)t^{1,1}t^{2,1}
-\frac{q}{1-q}t^{2,0}t^{2,1}\\
&+\log(1-q) \left(t^{1,1}\right)^2t^{2,1}
-\frac{2q}{1-q}t^{1,1}t^{2,0}t^{2,1}
-\frac{ q}{2(1-q)^2}\left(t^{2,0}\right)^2 t^{2,1}\\
&-\frac{q\log(1-q)}{1-q}\left(t^{2,1}\right)^3
+t^{1,0}\left(t^{1,1}\right)^2
-\frac{ q}{1-q}t^{2,0}\left(t^{2,1}\right)^2
+\cdots,\\
u^2_{\rm top,ad}
=&t^{2,0}
+t^{1,1}t^{2,0}
+t^{1,0}t^{2,1}
+\log(1-q)\left(t^{2,1}\right)^2
+2t^{1,0}t^{1,1}t^{2,1}\\
&+2\log(1-q)t^{1,1}\left(t^{2,1}\right)^2
+t^{2,0}\left(t^{1,1}\right)^2
-\frac{q}{1-q}t^{2,0}\left(t^{2,1}\right)^2
+\cdots.
\end{align*}

\section{Partial correlation functions}

The genus $g$ correlation functions
\beq
\llangle\tau_{i_1}(\phi_{\alpha_1})\cdots\tau_{i_k}(\phi_{\alpha_k})\rrangle_{g}
\eeq
evaluated at $t^{\alpha,p}=t_\alpha\delta^{p,0}$, $p\geq0$
are called the {\it partial correlation functions} (cf.~\cite{DYZ}),
and we still use the same notations to denote these partial correlation functions.

For the $\P$ model, we list the following genus $0,1,2$ partial correlation functions:
\begin{align*}
&\llangle\tau_1(\phi_1)\rrangle^{\P}_{0}
=\frac12t_1^2 t_2+q(t_2-2)e^{t_2},
\quad
\llangle\tau_1(\phi_2)\rrangle^{\P}_{0}
=\frac16(t_1)^3+qt_1e^{t_2}, 
\\
&\llangle\tau_1(\phi_1)\tau_1(\phi_1)\rrangle^{\P}_{0}
=(t_1)^2t_2+q\left((t_2)^2-2t_2+2\right)e^{t^2},
\\
&\llangle\tau_1(\phi_1)\tau_1(\phi_2)\rrangle^{\P}_{0}
=\frac13(t_1)^3+qt_1e^{t_2},
\quad
\llangle\tau_1(\phi_2)\tau_1(\phi_2)\rrangle^{\P}_{0}
=q(t_1)^2e^{t_2}+\frac12q^2e^{2t_2},
\\
&\llangle\tau_1(\phi_1)\rrangle^{\P}_{1}
=-\frac1{12}t_2+\frac1{12},
\quad
\llangle\tau_1(\phi_2)\rrangle^{\P}_{1}
=-\frac1{24}t_1, 
\\
&\llangle\tau_1(\phi_1)\tau_1(\phi_1)\rrangle^{\P}_{1}
=-\frac1{12}t_2+\frac1{12},
\quad
\llangle\tau_1(\phi_1)\tau_1(\phi_2)\rrangle^{\P}_{1}
=-\frac1{12}t_1,
\quad
\llangle\tau_1(\phi_2)\tau_1(\phi_2)\rrangle^{\P}_{1}
=0,\\
&\llangle\tau_1(\phi_\alpha)\rrangle_2^{\P}=\llangle\tau_1(\phi_\alpha)\tau_1(\phi_\beta)\rrangle_2^{\P}=0.
\end{align*}
Denote
\[
\omega:=\frac{qe^{t_2}}{1-qe^{t_2}},
\]
then for the resolved conifold $X$ with diagonal action,
we have
\begin{align*}
&\llangle\tau_1(\phi_1)\rrangle^{X,\rm di}_{0}
=-2\li_3(qe^{t_2})+t_2\li_2(qe^{t_2})+\frac13t_1^3+\frac12t_1^2t_2,
\\
&\llangle\tau_1(\phi_2)\rrangle^{X,\rm di}_{0}
=\left(\li_1(qe^{t_2})\right)^2+t_1\li_2(qe^{t_2})-t_2\li_1(qe^{t_2})+\frac16t_1^3,\\
&\llangle\tau_1(\phi_1)\tau_1(\phi_1)\rrangle^{X,\rm di}_{0}
=2\li_3(qe^{t_2})+\frac23t_1^3-2t_2\li_2(qe^{t_2})+t_1^2t_2+t_2^2\li_1(qe^{t_2}),\\
&\llangle\tau_1(\phi_1)\tau_1(\phi_2)\rrangle^{X,\rm di}_{0}
=-\left(\li_1(qe^{t_2})\right)^2+\frac13t_1^3+t_1t_2\li_1(qe^{t_2}),\\
&\llangle\tau_1(\phi_2)\tau_1(\phi_2)\rrangle^{X,\rm di}_{0}
=\int_{-\infty}^{t_2}\frac{(\log(1-qe^y))^2dy}{1-qe^y}-\left(\li_1(qe^{t_2})\right)^3-2t_2\left(\li_1(qe^{t_2}\right)^2+t_1^2\li_1(qe^{t_2}),\\
&\llangle\tau_1(\phi_1)\rrangle^{X,\rm di}_{1}
=\frac1{12}-t_2\left(\frac1{12}\omega-\frac1{24}\right), 
\\
&\llangle\tau_1(\phi_2)\rrangle_{1}^{X,\rm di}
=-\frac1{12}\omega+\left(\frac1{12}-\frac16\omega\right)\log(1+\omega)+t_1\left(-\frac1{24}+\frac1{12}\omega\right),
\\
&\llangle\tau_1(\phi_1)\tau_1(\phi_1)\rrangle^{X,\rm di}_{1}
=\frac1{12}+t_2\left(-\frac1{12}+\frac16\omega\right)+t_2^2\left(\frac1{12}\omega+\frac1{12}\omega^2\right),
\\
&\llangle\tau_1(\phi_1)\tau_1(\phi_2)\rrangle^{X,\rm di}_{1}\\
&\qquad
=-\frac1{12}\omega+\left(\frac1{12}-\frac16\omega\right)\log(1+\omega)
+t_1\left(-\frac1{12}+\frac16\omega\right)\\
&\qquad\quad+t_2\left(-\frac14\omega^2-\frac16\omega(1+\omega)\log(1+\omega)\right)
+\frac1{12}t_1t_2 \omega(1+\omega),
\\
&\llangle\tau_1(\phi_2)\tau_1(\phi_2)\rrangle^{X,\rm di}_{1} 
\\
&\qquad=\omega\left(\frac14+\frac16\omega\right)-\left(\frac1{12}-\frac16\omega-\omega^2\right)\log(1+\omega)
+\frac13\omega(1+\omega)\log(1-\omega)\\
&\qquad\quad+t_1\left(-\frac12\omega^2-\frac13\omega(1+\omega)\log(1+\omega)\right)
+\frac1{12}t_1^2\omega(1+\omega),\\
&\llangle\tau_1(\phi_1)\rrangle^{X,\rm di}_{2}
=\frac{-1+12\omega+12\omega^2}{1440}+\frac1{240}t_2\omega(1+\omega)(1+2\omega), 
\\
&\llangle\tau_1(\phi_2)\rrangle_{2}^{X,\rm di}
=-\frac{\omega^2(15+14\omega)}{1440}-\frac1{120}\omega(1+\omega)(1+2\omega)\log(1+\omega)
+\frac1{240}t_1\omega(1+\omega)(1+2\omega),
\\
&\llangle\tau_1(\phi_1)\tau_1(\phi_1)\rrangle^{X,\rm di}_{2}
=\frac{-1+12\omega+12\omega^2}{480}+\frac1{40}t_2\omega(1+\omega)(1+2\omega)
+\frac1{240}t_2^2\omega(1+\omega)(1+6\omega+6\omega^2),
\\
&\llangle\tau_1(\phi_1)\tau_1(\phi_2)\rrangle^{X,\rm di}_{2}\\
&\qquad
=-\frac{\omega^2(15+14\omega)}{480}-\frac1{40}\omega(1+\omega)(1+2\omega)\log(1+\omega)
+\frac1{60}t_1\omega(1+\omega)(1+2\omega)\\
&\qquad\quad-\frac1{240}t_2\omega(1+\omega)(\omega(7+11\omega)+2(1+6\omega+6\omega^2))
+\frac1{240}t_1t_2\omega(1+\omega)(1+6\omega+6\omega^2),
\\
&\llangle\tau_1(\phi_2)\tau_1(\phi_2)\rrangle^{X,\rm di}_{2} 
\\
&\qquad=-\frac{\omega^2(61+74\omega+16\omega^2)}{1440}
+\frac{\omega(1+\omega)(1+16\omega+22\omega^2)\log(1+\omega)}{120}\\
&\qquad\quad+\frac{\omega(1+\omega)(1+6\omega+6\omega^2)(\log(1+\omega))^2}{60}+\frac1{240}t_1^2\omega(1+\omega)(1+6\omega+6\omega^2)\\
&\qquad\quad-\frac1{120}t_1\omega(1+\omega)\left(\omega(7+11\omega)+2(1+6\omega+6\omega^2)\log(1+\omega)\right).
\end{align*}
For the resolved conifold $X$ with anti-diagonal action,
we have
\begin{align*}
&\llangle\tau_1(\phi_1)\rrangle^{X,\rm ad}_{0}
=\frac12t_1^2t_2+2\li_3(qe^{t_2})-t_2\li_2(qe^{t_2}), 
\quad
\llangle\tau_2(\phi_1)\rrangle_{0}^{X,\rm ad}
=\frac16t_1^3-t_1\li_2(qe^{t_2}),
\\
&\llangle\tau_1(\phi_1)\tau_1(\phi_1)\rrangle^{X,\rm ad}_{0}
=-2\li_3(qe^{t_2})+2t_2\li_2(qe^{t_2})+t_1^2t_2-t_2^2\li_1(qe^{t_2}),
\\
&\llangle\tau_1(\phi_1)\tau_1(\phi_2)\rrangle^{X,\rm ad}_{0}
=\frac13t_1^3-t_1t_2\li_1(qe^{t_2}),
\quad
\llangle\tau_1(\phi_2)\tau_1(\phi_2)\rrangle^{X,\rm ad}_{0}
=-t_1^2\li_1(qe^{t_2})+\int_{-\infty}^{t_2}\left(\li_1(qe^y)\right)^2dy,
\\
&\llangle\tau_1(\phi_1)\rrangle^{X,\rm ad}_{1}
=\frac1{12}+t_2\left(-\frac1{24}+\frac1{12}\omega\right), 
\quad
\llangle\tau_2(\phi_1)\rrangle_{1}^{X,\rm ad}
=t_1\left(-\frac1{24}+\frac1{12}\omega\right),
\\
&\llangle\tau_1(\phi_1)\tau_1(\phi_1)\rrangle^{X,\rm ad}_{1}
=\frac1{12}+t_2\left(-\frac1{12}+\frac16\omega\right)+\frac1{12}t_2^2\omega(1+\omega),
\\
&\llangle\tau_1(\phi_1)\tau_1(\phi_2)\rrangle^{X,\rm ad}_{1}
=t_1\left(-\frac1{12}+\frac16\omega\right)+\frac1{12}t_1t_2\omega(1+\omega),
\\
&\llangle\tau_1(\phi_2)\tau_1(\phi_2)\rrangle^{X,\rm ad}_{1}
=-\frac{\omega(1+2\omega)}{12(1+\omega)}+\frac1{12}(1-2\omega)\log(1+\omega)+\frac1{12}t_1^2\omega(1+\omega),\\
&\llangle\tau_1(\phi_1)\rrangle^{X,\rm ad}_{2}
=\frac{1+12\omega+12\omega^2}{1440}+\frac1{240}t_2\omega(1+\omega)(1+2\omega), 
\\
&\llangle\tau_2(\phi_1)\rrangle_{2}^{X,\rm ad}
=\frac1{240}t_1\omega(1+\omega)(1+2\omega),
\\
&\llangle\tau_1(\phi_1)\tau_1(\phi_1)\rrangle^{X,\rm ad}_{2}
=\frac{1+12\omega+12\omega^2}{480}+\frac1{40}t_2\omega(1+\omega)(1+2\omega)
+\frac1{240}t_2^2\omega(1+\omega)(1+6\omega+6\omega^2),
\\
&\llangle\tau_1(\phi_1)\tau_1(\phi_2)\rrangle^{X,\rm ad}_{2}
=\frac1{60}t_1\omega(1+\omega)(1+2\omega)+\frac1{240}t_1t_2\omega(1+\omega)(1+6\omega+6\omega^2),
\\
&\llangle\tau_1(\phi_2)\tau_1(\phi_2)\rrangle^{X,\rm ad}_{2}
=-\frac{\omega^2(59+62\omega)}{1440}-\frac1{120}\omega(1+\omega)(1+2\omega)\log(1+\omega)
+\frac1{240}t_1^2\omega(1+\omega)(1+6\omega+6\omega^2).
\end{align*}
	
\end{appendices}

\begin{center}
{\small
Si-Qi Liu, Department of Mathematical Sciences, Tsinghua University, Beijing 100084, P.R.~China\\ 
e-mail: liusq@mail.tsinghua.edu.cn\\
~\\
Di Yang, School of Mathematical Sciences, University of Science and Technology of China,\\
Hefei 230026, P.R.~China\\
e-mail: diyang@ustc.edu.cn\\
~\\
Youjin Zhang, Department of Mathematical Sciences, Tsinghua University, Beijing 100084, P.R.~China\\
e-mail: youjin@mail.tsinghua.edu.cn\\
~\\
Chunhui Zhou, Institute of Geometry and Physics, University of Science and Technology of China,\\ 
Hefei 230026, P.R.~China\\
e-mail: zhouch@ustc.edu.cn}
\end{center}


\begin{thebibliography}{99}
	
\bibitem{BP} Belorousski, P., Pandharipande, R.,  A descendent relation in genus 2. 
Ann. Scuola Norm. Sup. Pisa Cl. Sci. (4) {\bf 29} (2000), 171--191.

\bibitem{BF}
Behrend K., Fantechi B., 
The intrinsic normal cone, Invent. Math. {\bf 128} (1997), 45--88.

\bibitem{Bri}
Brini A., The local Gromov-Witten theory of $\P$ and integrable hierarchies,
Comm. Math. Phys. {\bf 313} (2012), 571--605.

\bibitem{BCR}
Brini A., Carlet G., Rossi P., Integrable hierarchies and the mirror model of local $\P$,
Phys. D, {\bf 241} (2012), 2156--2167.

\bibitem{BrP}
Bryan, J., Pandharipande, R., 
The local Gromov-Witten theory of curves, 
J. Amer. Math. Soc. {\bf 21} (2008), 101--136.

\bibitem{CDZ}
Carlet G., Dubrovin B., Zhang Y., The extended Toda hierarchy,
Mosc. Math. J. {\bf 4} (2004), 313--332.

\bibitem{DW}
Dijkgraaf R.,  Witten E., 
Mean field theory, topological field theory, and multi-matrix models, Nucl. Phys. B {\bf 342} (1990), 486--522.

\bibitem{Du0}
Dubrovin, B., 
Integrable systems and classification of 2D topological field theories, 
in ``Integrable Systems", The J.-L.Verdier Memorial Conference, Actes du Colloque International de Luminy, 
Babelon, O., Cartier, P., Kosmann-Schwarzbach, Y. (eds.), 
 pp. 313--359. Birkh\"auser, 1993.

\bibitem{Du1}
Dubrovin B., Geometry of 2D topological field theories, 
in ``Integrable Systems and Quantum Groups" (Montecatini Terme, 1993),
Francaviglia M., Greco S. (eds.), 
Springer Lecture Notes in Math. {\bf 1620}, pp. 120--348, 1996.

\bibitem{Du2}
Dubrovin B., On almost duality for Frobenius manifolds,
in ``Geometry, topology, and mathematical physics", Amer. Math. Soc. Transl. Ser. (2) {\bf 212}, pp. 75--132, Amer. Math. Soc., Providence, RI, 2004.

\bibitem{Du3}
Dubrovin B., On universality of critical behaviour in Hamiltonian PDEs,
Amer. Math. Soc. Transl. {\bf 224} (2008), 59--109.

\bibitem{Du4}
Dubrovin, B., 
Hamiltonian perturbations of hyperbolic PDEs: from classification results to the properties of solutions, 
in ``New Trends in Mathematical Physics'', Sidoravi\v cius V. (ed.), 
pp. 231--276, Springer, Dordrecht, 2009.

\bibitem{DLYZ0}
Dubrovin, B., Liu, S.-Q., Yang, D., Zhang, Y., 
Hodge integrals and tau-symmetric integrable hierarchies of Hamiltonian evolutionary PDEs. Adv. Math. {\bf 293} (2016), 382--435.

\bibitem{DLYZ}
Dubrovin B., Liu S.-Q., Yang D., Zhang Y., 
Hodge-GUE correspondence and the discrete KdV equation, Comm. Math. Phys. {\bf 379} (2020), 461--490.

\bibitem{DLZ}
Dubrovin B., Liu S.-Q., Zhang Y., 
On the genus two free energies for semisimple Frobenius manifold,
Russ. J. Math. Phys. {\bf 19} (2012), 273--298.

\bibitem{DY1}
Dubrovin B., Yang D., 
Generating series for GUE correlators,
Lett. Math. Phys. {\bf 107} (2017), 1971--2012.

\bibitem{DY2}
Dubrovin B., Yang D., 
On cubic Hodge integrals and random matrices,
Comm. Number Theory Phys. {\bf 11} (2017), 311--336.

\bibitem{DYZ}
Dubrovin B., Yang D., Zagier D., 
Gromov--Witten invariants of the Riemann sphere,
Pure Appl. Math. Q. {\bf 16} (2020), 153--190.

\bibitem{DZ2}
Dubrovin B., Zhang Y.,
Frobenius manifold and Virasoro constraints, 
Selecta Math. 
{\bf 5} (1998),
423--466.

\bibitem{DZ3}
Dubrovin B., Zhang Y.,
Bi-hamiltonian hierarchies in 2D topological field theory at one-loop approximation,
Comm. Math. Phys. {\bf 198} (1998), 311--361.

\bibitem{DZ-norm}
Dubrovin, B., Zhang, Y., Normal forms of hierarchies of integrable PDEs, Frobenius manifolds and Gromov--Witten invariants, arXiv:math/0108160.

\bibitem{DZ1}
Dubrovin B., Zhang Y., Virasoro symmetries of the extended Toda hierarchy, 
Comm. Math. Phys. {\bf 250} (2004), 161--193.

\bibitem{EGX}
Eguchi T., Getzler E., Xiong C.-S., 
Topological gravity in genus 2 with two primary fields,
Adv. Theor. Math. Phys. {\bf 4} (2000), 981--1000.

\bibitem{EYY}
Eguchi T., Yamada Y., Yang S.-K.,
On the genus expansion in the topological string theory,
Rev. Math. Phys. {\bf 7} (1995), 279--309. 

\bibitem{EY}
Eguchi, T., Yang, S.-K., The topological CP1 model and the large-N matrix integral. Modern Physics Letters A~{\bf 9} (1994), 2893--2902.

\bibitem{Getz} 
Getzler E.,
The Toda conjecture, 
in ``Symplectic geometry and mirror symmetry" (Seoul, 2000), pp. 51--79, World Sci. Publishing, River Edge, NJ, 2001.

\bibitem{G}
Getzler E.,
Intersection theory on $\overline{\mathcal M}_{1,4}$ and elliptic Gromov--Witten invariants, 
J. Amer. Math. Soc. {\bf 10} (1997), 973--998.

\bibitem{G2}
Getzler E., 
Topological recursion relations in genus 2,
in ``Integrable systems and algebraic geometry" (Kobe/Kyoto, 1997), pp. 73--106,
World Sci. Publishing, River Edge, NJ, 1998.

\bibitem{KM}
Kontsevich M., Manin Yu., 
Gromov--Witten classes, quantum cohomology and enumerative geometry, Comm. Math. Phys. {\bf 164} (1994), 525--562.

\bibitem{LT}
Li J., Tian G., 
Virtual moduli cycles and Gromov--Witten invariants of algebraic varieties, J. Amer. Math. Soc. {\bf 11} (1998), 119--174.

\bibitem{LYZZ}
Liu S.-Q., Yang D., Zhang Y., Zhou C., 
The Hodge-FVH correspondence,
J. Reine Angew. Math. {\bf 775} (2021), 259--300.

\bibitem{Mum}
Mumford A., Towards an enumerative geometry of the moduli space of curves, 
in ``Arithmetic and geometry", vol. II, Progr. Math., {\bf 36}, pp. 271--328, Birkh\"auser, Boston, 1983.

\bibitem{OP}
Okounkov A., Pandharipande R., Gromov--Witten theory, Hurwitz theory and completed cycles, 
Ann. of Math. {\bf 163} (2006), 517--560.

\bibitem{OP1}
Okounkov A., Pandharipande R.,
The equivariant Gromov--Witten theory of $\mathbb P^1$, Ann. of Math. {\bf 163} (2006), 561--605.

\bibitem{RT1}
Ruan Y., Tian G., A mathematical theory of quantum cohomology, J. Diff. Geom. {\bf 42} (1995), 259--367.

\bibitem{V1}
Vekslerchik V.~E., Functional representation of the Ablowitz--Ladik hierarchy, J. Phys. A {\bf 31} (1998), 1087--1099.

\bibitem{V2}
Vekslerchik V.~E., Universality of the Ablowitz--Ladik hierarchy, arXiv:9807005.

\bibitem{W}
Witten E., 
Two dimensional gravity and intersection theory on moduli space, Surveys in Diff. Geom. {\bf 1} (1991), 243--310.

\bibitem{Z}
Zhang Y.,
On the $CP^1$ topological sigma model and the Toda lattice hierarchy,
J. Geom. Phys. {\bf 40} (2002), 215--232.

\end{thebibliography}
\end{document}